\documentclass[journal]{IEEEtran}
\usepackage[caption=false,font=normalsize,labelfont=sf,textfont=sf]{subfig}
\usepackage{cite}
\usepackage{graphics}
\usepackage{amsmath}		
\usepackage{amsfonts}		
\usepackage{amssymb}        
\usepackage{ulem}
\usepackage{svg}
\ifCLASSINFOpdf
\usepackage{graphicx}
\usepackage{txfonts}
\else
\usepackage{graphicx}
\fi
\usepackage{float}
\usepackage{tabu}
\usepackage{booktabs}
\usepackage{placeins}
\usepackage{multicol}
\usepackage{epstopdf}
\usepackage{breqn}
\graphicspath{ {./Images/} }
\usepackage{multirow}
\usepackage{hyperref}
\usepackage{caption}
\usepackage{soul}
\usepackage{algorithm}
\usepackage{algpseudocode}
\usepackage{mathtools}
\usepackage{xcolor}

\usepackage{bbm}
\usepackage{dblfloatfix}

\usepackage{hyperref}
\hypersetup{
    colorlinks=true,
    linkcolor=blue,
    filecolor=magenta,      
    urlcolor=blue,
}

\definecolor{amber(sae/ece)}{rgb}{1.0, 0.49, 0.0}
\definecolor{aqua}{rgb}{0.0, 1.0, 1.0}

\begin{document}

\title{Low Complexity Deep Learning Augmented Wireless Channel Estimation for Pilot-Based OFDM on Zynq System on Chip}

\author{Animesh Sharma*, Syed Asrar Ul Haq*,  and Sumit J. Darak, \textit{Senior Member, IEEE}
  \thanks{*Animesh Sharma, and Syed Asrar Ul Haq are joint first authors.}
					\thanks{This work is supported by the CHANAKYA PG fellowship awarded to Animesh Sharma from DRISHTI CPS foundation, IIT Indore and funding received from core research grant (CRG) awarded to Dr. Sumit J. Darak from DST-SERB, GoI.}

			\thanks{Animesh Sharma, Syed Asrar Ul Haq and Sumit J. Darak are with Electronics and Communications Department, 
				IIIT-Delhi, India-110020 (e-mail: \{animesh20317,syedh,sumit\}@iiitd.ac.in}
			
}
	\maketitle
\begin{abstract}

Channel estimation (CE) is one of the critical signal-processing tasks of the wireless physical layer (PHY). Recent deep learning (DL) based CE have outperformed statistical approaches such as least-square-based CE (LS) and linear minimum mean square error-based CE (LMMSE). However, existing CE approaches have not yet been realized on system-on-chip (SoC). The first contribution of this paper is to efficiently implement the existing state-of-the-art CE algorithms on Zynq SoC (ZSoC), comprising of ARM processor and field programmable gate array (FPGA), via hardware-software co-design and fixed point analysis. We validate the superiority of DL-based CE and LMMSE over LS for various signal-to-noise ratios (SNR) and wireless channels in terms of mean square error (MSE) and bit error rate (BER). We also highlight the high complexity, execution time, and power consumption of DL-based CE and LMMSE approaches. To address this, we propose a novel compute-efficient LS-augmented interpolated deep neural network (LSiDNN) based CE algorithm and realize it on ZSoC. 
The proposed LSiDNN offers 88-90\% lower execution time and 38-85\% lower resource utilization than state-of-the-art DL-based CE for identical MSE and BER. LSiDNN offers significantly lower MSE and BER than LMMSE, and the gain improves with increased mobility between transceivers. It offers 75\% lower execution time and 90-94\% lower resource utilization than LMMSE.
           
\end{abstract}

\newcommand{\pdob}{$P_d^{OB}$} 
\newcommand{\pdab}{$P_d^{AB}$} 

\section{Introduction}
\label{Sec:Intr}
 The evolution of wireless PHY from 1G to 5G has seen significant improvement in its intelligence and reconfiguration capability via on-the-fly selection of modulation and coding scheme, waveform type, carrier frequency, and bandwidth. The emergence of next-generation vehicular, high-speed multimedia, and security applications demand high throughput and ultra-reliable low-latency PHY \cite{Evolution}. In this direction, various innovations such as wider bandwidth, massive multiple input multiple outputs (massive MIMO), beamforming, and millimeter-wave systems are being explored. In addition, recent advances in machine learning (ML) and deep learning (DL) have been shown to offer improved performance \cite{DLforPhy2017oShea, DLRohith,DLBasedReceiver2017, end2endGAN,ComNet2018 , CEandSD2018,DeepReceiver2021,DeepWiPHY2021,Gizzini2020STADNN,HOU2023GRUbasedCE,  Gizzini2020trfidnn, han2019AEDNN, 2021GizziniLstm,revMimo,revDetection,sigDetection, haq2022LSDNN,vlsidLsdnn,channelNet,cite:ReEsNet,iResNet,transformerBasedCE,HA02,Channelformer, FreqTimeNet}. The design of ML and DL-based PHY is an exciting research direction, with various open research questions such as relevance of statistical algorithms, reliability in dynamic environments, compatibility with existing standards, and feasibility on hardware platforms.

Recent works have explored DL for end-to-end design of wireless PHY \cite{DLforPhy2017oShea, DLBasedReceiver2017, end2endGAN,DLRohith}. However, such DL models are computationally complex with a large number of parameters, making them difficult to map on hardware edge platforms used in wireless base stations and access points. Since DL-based approaches are dataset-dependent, they must be configured whenever channel statistics change. The reconfiguration of large-size DL models is time and memory-intensive. The alternative approach is the block-based DL approach, where one or more signal processing blocks are replaced by a DL model \cite{ComNet2018 , CEandSD2018, DeepWiPHY2021, DeepReceiver2021}.   ComNet \cite{ComNet2018} uses two neural networks to replace the channel estimation and signal detection blocks in an OFDM receiver chain. It incorporates expert knowledge by taking the output of conventional channel estimation and zero-forcing equalizer block as input to respective deep neural networks (DNNs). \cite{CEandSD2018} demonstrates the direct recovery of transmitted bits from the received signal using a single DNN, replacing demodulation, channel estimation, and signal detection blocks. DeepReceiver \cite{DeepReceiver2021} includes a synchronization block in its DNN model, along with the CE and signal detection blocks, demonstrating that DNNs can combat non-linear effects and exhibit anti-jamming properties. DeepWiPHY \cite{DeepWiPHY2021} replaces channel estimation, equalization, demapping, and synchronization modules using two DNN models trained together to obtain constellation points directly from the received signals. The block-based approach ~\cite{Gizzini2020STADNN,HOU2023GRUbasedCE, Gizzini2020trfidnn, han2019AEDNN, 2021GizziniLstm,revMimo} is standard-compatible since the intermediate feedback, such as channel quality indicator (CQI), precoder matrix indicator (PMI), and rank indicator (RI) signals, can be estimated as desired in existing standards. Various other works~\cite{revDetection,sigDetection} have jointly explored DL-based CE and signal detection. However, such approaches can not provide CQI, which is crucial for selecting PHY parameters by the transmitter for subsequent communication and hence, incompatible with existing standards.

In the June 2021 3GPP workshop, various industry leaders presented the framework for the evolution of intelligent and reconfigurable wireless PHY, emphasizing the commercial potential of AI/ML in the wireless domain. 3GPP has included study and work items dedicated to AI/ML applications in wireless networks in release 18. These study items span from the application layer to the AI-native air interface, covering a broad range of AI/ML applications for the advanced 5G network~\cite{3gppStudyItems}. In this direction, various industry leaders have recently demonstrated DL-based wireless PHY in a real radio environment. DeepSig demonstrated OmniSIG, an end-to-end DL-based wireless communication system obtained by replacing the transmitter-receiver chain with an autoencoder model~\cite{omnisig}. Nokia Bell Labs introduced DeepRX~\cite{DeepRx}, which replaces CE, equalization, and demodulation with a DL model. Their experiments in a 5G system showed improved performance compared to the benchmark LMMSE receiver. In collaboration with Rohde and Schwarz, Nvidia demonstrated the Neural Receiver~\cite{neuralreceiver} for a multi-user multi-antenna system. They showed that DL-based PHY offers comparable performance to MMSE-based PHY with significant savings in complexity. These recent works highlight significant interests and feasibility of AI/ML/DL for wireless PHY in academia and industry.




CE is one of the critical signal-processing tasks of the wireless PHY. 
Historically, the CE for wireless PHY can be classified into two broad categories: 1) Preamble-based CE in IEEE 802.11 standards, and 2) Pilot-based CE in cellular standards. In preamble-based wireless PHY, the CE over all the sub-carriers of orthogonal frequency division multiplexing (OFDM) PHY is done using the preamble transmitted at the beginning of the data frame. In a recent standard, IEEE 802.11bd, midamble is transmitted at the middle of the data frame to improve the CE. The pilot-based CE in cellular PHY is more complex since the pilots are transmitted only over a few sub-carriers. Hence, two-dimensional (2D) interpolation is needed to obtain the CE at all sub-carriers. Statistical least-squares (LS) and linear minimum mean square estimation (LMMSE) are widely used for CE in both cases. With new services such as data-intensive multimedia, ultra-reliable low-latency communication, and vehicular communication, wireless PHY with wider bandwidth and fewer pilots are desired. The presence of a dynamic mobile channel environment further increases the computational complexity of the CE.  In these applications, LS has shown poor performance, particularly in low SNR conditions, while LMMSE requires prior knowledge of wireless channel parameters. The fixed-point realization of LMMSE poses challenges due to inverse matrix operations resulting in a significant area, delay, and power overheads\cite{haq2022LSDNN}. Hence, DL-based CE needs to be explored. Our previous work \cite{haq2022LSDNN,vlsidLsdnn} shows that DL-augmented CE offers superior performance over LS and LMMSE for preamble-based PHY. In this work, we focus on pilot-based PHY.

      \begin{table*}[!b]

\centering
\caption{\small Comparison of existing DL-based channel estimation approaches for pilot-based OFDM PHY. }
\label{tab:existingComparison}
\renewcommand{\arraystretch}{1.2}
 \resizebox{\textwidth}{!}
 {
\begin{tabular}{|l|l|l|l|l|l|l|l|}
\hline
\textbf{Approach}             & \textbf{DL model} & \textbf{Standard Compatible} & \textbf{Computational Complexity} & \textbf{Memory Complexity} & \textbf{Latency} & \textbf{HW IP}  & \textbf{Fixed-point} \\ \hline
\textbf{\cite{channelNet}}  & CNN   & Yes   & Very High & Very High & Very High & No  & No  \\ \hline
\textbf{\cite{cite:ReEsNet}}    & RNN   & Yes   & High  & Medium    & Medium    & No & No   \\ \hline
\textbf{\cite{iResNet}}     & RNN   & Yes   & Medium    & Low   & Medium    & No  & No  \\ \hline

\textbf{\cite{transformerBasedCE}}  & Transformer, Conv & Yes   & Very High    & Medium    & Medium    & No  & No  \\ \hline
\textbf{\cite{HA02}}    & Transformer, RNN  & Yes   & Very High    & High  & High  & No  & No  \\ \hline
\textbf{\cite{FreqTimeNet}} & FCNN, Attention   & Yes   & Very High    & High  & Medium    & No & No\\ \hline
\textbf{Proposed}   & FCNN  & Yes   & Low   & High  & Low   & Yes & Yes   \\ \hline
\end{tabular}
}
\end{table*}

Convolutional neural networks (CNNs) based CE exploiting the 2D nature of the channel matrix has been discussed in \cite{channelNet,cite:ReEsNet,iResNet}. ChannelNet in  \cite{channelNet} employs two cascaded CNNs to convert the channel matrix, represented as a low-resolution image on pilot locations, to a high-resolution image of noisy channel estimates at all the sub-carriers.
However, it suffers from high computational complexity, large memory requirements, and high latency due to 23 layer architecture. 
An improved low complexity CE in \cite{cite:ReEsNet} is based on a residual learning neural network (ReEsNet) instead of two large CNNs. The interpolated ReSNet (IResNet) in \cite{iResNet} further reduces the computational complexity by replacing the transposed convolution in ResNet with a bilinear interpolation layer. This approach also makes it compatible with flexible pilot patterns. 

Recent advances in DL, such as transformers and attention-based approaches, have been discussed for CE. \cite{transformerBasedCE} utilizes a single transformer module in combination with a CNN, while \cite{HA02, Channelformer} incorporates a ResNeT and an upsampling network into a transformer module. The transformer is an encoder to extract features from LS estimates, while a residual network module is utilized for CE at pilot locations. An upsampling module, consisting of a fully connected layer and a CNN layer, is employed to interpolate the estimates from pilot locations to the entire time-frequency grid.
FreqTimeNet \cite{FreqTimeNet} is another approach that divides the OFDM time-frequency grid into separate time and frequency components. It employs a fully connected neural network (FCNN) to obtain initial channel estimates and interpolate them along the subcarriers. 
Building upon FreqTimeNet, AttenFreqTimeNet \cite{FreqTimeNet} introduces an attention module incorporating SNR information in the estimation, making it more robust against SNR variations.

DL-based CE is critical in complex electromagnetic conditions such as mining~\cite{mineOfdm}, underwater acoustic systems~\cite{UWA}, and new frequency spectrum (6 GHz - 52 GHz). In these environments, characterized by rough and irregular surfaces or the absence of suitable mathematical models, DL-based channel estimation emerges as a promising alternative. In such situations, channel characteristics can be directly extracted from the data without relying on mathematical modeling. The DL model can be trained using high-complexity MMSE estimation as labels. Another approach involves using high SNR reference signals as labels in real-world signals~\cite{Channelformer}. Passive sniffing is also a potential approach to capture the training data~\cite{DeepWiPHY2021}. The combination of simulation and real-world data can be utilized to train the DL model for CE.

Most of the existing DL-based CE replaces the conventional statistical LS and LMMSE approaches completely. This results in significantly high complexity, power consumption, and low throughput. Furthermore, none of the existing DL-based CE for pilot-based OFDM have been realized on edge platforms such as system-on-chip (SoC). The proposed work aims to address these research gaps with innovative solutions at the algorithm and architecture levels with a system-level demonstration on the SoC. 
From an algorithm perspective, we propose a novel LS-augmented DL approach instead of replacing the LS or LMMSE completely with DL. From an architecture perspective, we develop software and hardware IPs for existing statistical approaches, state-of-the-art DL approaches, and proposed approaches for the demonstration on the Zynq SoC (ZSoC) from AMD-Xilinx. 
The GitHub repository containing simulation codes and hardware IPs can be accessed at~\cite{ githubRepo}.
For easier understanding, we compare the novel contributions of this paper to existing works in Table \ref{tab:existingComparison}. The contributions of this paper can be summarised below.

\begin{enumerate}
    \item Existing statistical and DL-based CE approaches for pilot-based OFDM PHY are mapped on the ZSoC via hardware-software co-design (HSCD) and fixed-point analysis. We optimize the state-of-the-art IResNet in \cite{iResNet} to improve its resource, execution time, and power performance without compromising functional accuracy. We highlight the high resource utilization, execution time, and power consumption of state-of-the-art CE architectures. 
    \item We develop a novel LS-augmented interpolated DNN-based CE (LSiDNN), analyze its performance and efficiently map it on the SoC via HSCD and fixed point analysis.
  We demonstrate substantial savings in computational complexity and execution time without significant degradation in functional accuracy for a wide range of signal-to-noise ratio (SNR) and wireless channels. Specifically, the proposed LSiDNN approach offers 88-90\% lower execution time and 38-85\% lower resources than recent DL-based CE. 
    \item DL-based CE offers superior performance than LMMSE for a wide range of Doppler velocities without any need for retraining. LSiDNN significantly outperforms LMMSE in CE accuracy. It also offers 75\% lower execution time, 30\% lower power consumption, and 90-94\% lower resource utilization than the LMMSE.
    
    \item     The proposed memory-based adaptable architecture enables DL-based CE to support different channels using the same hardware as long as corresponding DL parameters are stored in memory.
    
\end{enumerate}

The rest of the paper is organized as follows. We discuss the literature review in section II, the System model in section III, Channel Estimation approaches in section IV, Performance Analysis and Complexity Comparison in section V, and References in section VI.

\section{System Model}\label{Sec:SM}

\begin{figure}[!t]
    \centering
    \includegraphics[scale=0.575]{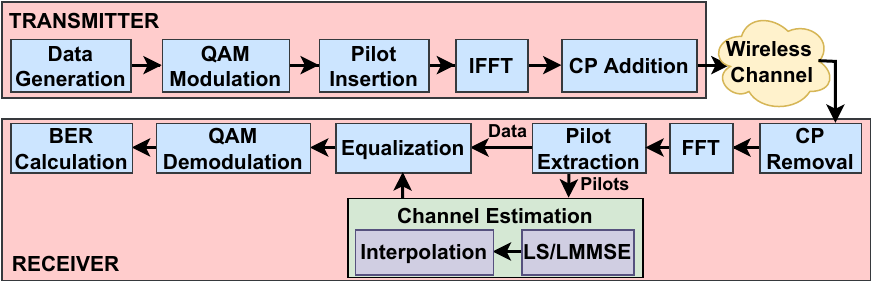}
    \caption{Block diagram of an OFDM-based transceiver PHY. }
    \label{fig:blockDiagram}
\end{figure}

\begin{figure}[!t]
\vspace{-0.2cm}
    \centering
    \includegraphics[scale=0.625]{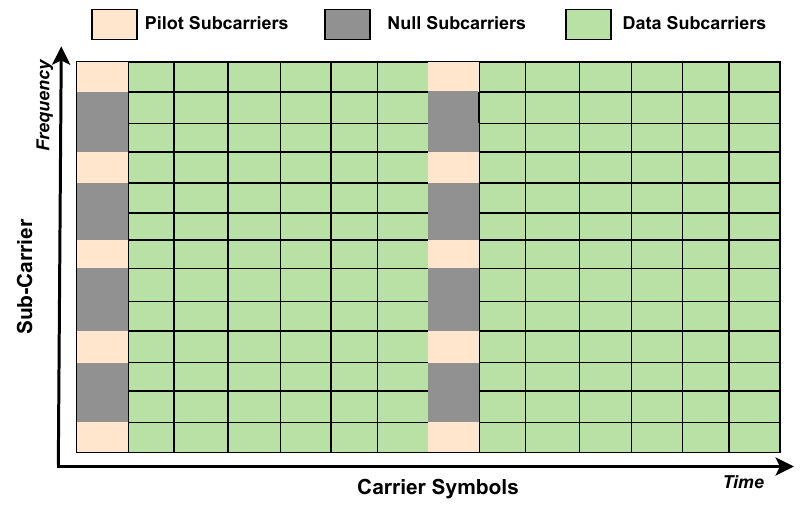}
    \caption{OFDM frame consisting of pilot and data sub-carriers.}
    \label{fig:frameStruct}
\end{figure}

We consider OFDM-based wireless PHY with a comb-pilot pattern as per the 3GPP standards. The transceiver PHY building blocks are shown in Fig.~\ref{fig:blockDiagram}. Each data frame consists of $N_s$ number of OFDM symbols and $N_{f}$ number of subcarriers in each OFDM symbol, as shown in Fig.~\ref{fig:frameStruct}. Based on the applications and 3GPP standards, the frame comprises a certain number of pilot, null, and data sub-carriers. Similar to \cite{cite:ReEsNet} \cite{iResNet}, we assume 72 sub-carriers per OFDM symbols. In a frame, pilots are transmitted over the first and seventh OFDM symbols, and there are 24 pilot sub-carriers per OFDM symbols interleaved by null sub-carriers. All the existing and proposed CE algorithms discussed in this paper can be easily adapted to other pilot and frame patterns.

The data and pilot sub-carriers undergo the data and OFDM modulation as shown in Fig.~\ref{fig:blockDiagram}. At the receiver, we assume ideal timing and phase synchronization. OFDM demodulation is performed, and pilot sub-carriers are extracted for CE. This is followed by channel equalization over the received data frame. In the end, data is demodulated and demapped to recover the received bits. We use normalized mean square error (NMSE)  and bit-error-rate (BER) as performance metrics for CE and end-to-end transceivers.



When a transmitted signal traverses a wireless channel, it undergoes fading caused by multipath effects and Doppler shifts due to mobility. These phenomena collectively affect the channel gain experienced by each subcarrier in an OFDM system. CE outputs the channel gain matrix using a predefined known signal, i.e., pilots. Since we can not transmit pilots over all subcarriers and the number of pilot sub-carriers should be as small as possible for higher throughput and spectrum efficiency, we calculate the channel gain at the pilot sub-carriers and estimate the same at the rest of the sub-carriers of the data frame. In OFDM systems, the frequency domain input-output relationship of the ${i^{th}}$ symbol and the ${k^{th}}$ sub-carrier is represented as:
\begin{align}
    \mathbf{Y}_{i,k} =\mathbf{H}_{i,k}\mathbf{X}_{i,k} + \mathbf{Z}_{i,k}; 
\end{align}\normalsize
where, ${Y_{i,k}}$ corresponds to the received signal, ${X_{i,k}}$ and ${Z_{i,k}}$ corresponds to transmitted data and Additive White Gaussian Noise, respectively. ${H_{i,k}}$ represents channel matrix at ${i^{th}}$ symbol and ${k^{th}}$ sub-carrier.  


  The LS and LMMSE are state-of-the-art statistical CE approaches. The LS minimizes the squared difference between the received symbol and the channel response \cite{LS+MMSE_literature}. Its output at any pilot position $p$ is given by
    \begin{align}
        \mathbf{H}^{LS}_p = min(||\mathbf{Y}_{p} - \mathbf{H}_{p}\mathbf{X}_{p}||^{2})
        \label{formula:LS}
    \end{align}\normalsize
    where, ${Y}_{p}$, ${H}_{p}$ and ${X_{p}}$ are the received channel response, channel gain, and transmitted signal, respectively, at the pilot $p$. For OFDM PHY, LS can be simplified as \cite{MIMOBook}
    \begin{align}
        \mathbf{H}^{LS}_p = \mathbf{Y}_{p} / \mathbf{X}_{p}
        \label{formula:LS_final}
    \end{align}\normalsize

\begin{figure}[!b]
    \centering
     \includegraphics[scale=0.85]{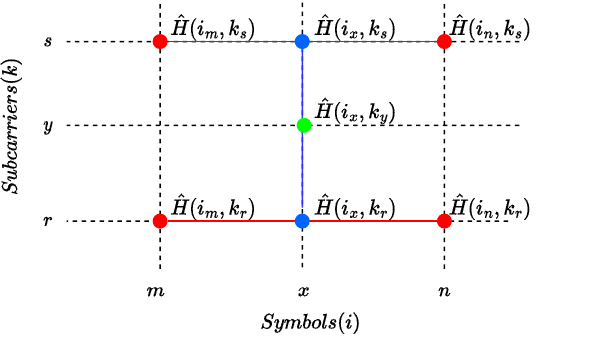}
    \caption{Illustrative example of bilinear interpolation.}
    \label{fig:Bilinear Interpolation}
\end{figure}

    After estimating the channel gain for all pilot sub-carriers and pilot symbols, the channel gain for the data subcarriers across the entire frame is obtained through Bilinear interpolation. Let \(i_m\) and \(i_n\) represent the pilot symbol indices, and \(k_r\) and \(k_s\) represent subcarrier indices for pilot subcarriers within pilot symbols, as illustrated in figure~\ref{fig:Bilinear Interpolation}. To obtain the interpolated channel estimates \(\hat{H}(i,k)\), for data symbols, we start by interpolating the estimates between pilot symbols along the time axis. Then the esimates along frequency axis are calculated.  
For a data symbol at the symbol index \(i_x\), the interpolated estimates at pilot subcarrier indices are given by:
\begin{equation}
    \hat{H}(i_x,k_r) = \frac{i_n-i_x}{i_n-i_m}\times \hat{H}(i_m,k_r)+\frac{i_x-i_m}{i_n-i_m}\times \hat{H}(i_n,k_r) \notag
\end{equation}
\begin{equation}
    \hat{H}(i_x,k_s) = \frac{i_n-i_x}{i_n-i_m}\times \hat{H}(i_m,k_s)+\frac{i_x-i_m}{i_n-i_m}\times \hat{H}(i_n,k_s)  \notag
\end{equation}

These interpolated estimates are used to obtain estimate for data subcarrier corresponding to non-pilot subcarrier index \(i_y\) as follows:
\begin{equation}
     \hat{H}(i_x,k_y) = \frac{k_s-k_y}{k_s-k_r}\times \hat{H}(i_x,k_r) + \frac{k_y-k_r}{k_s-k_r}\times \hat{H}(i_x,k_s)
\end{equation}

    The LS-based CE does not need prior knowledge of the channel statistics or noise. Moreover, considering that pilot values are typically constant, it involves a complex division operation between a constant complex value and the received complex signal value, making it hardware-friendly.

    The LMMSE based CE is an enhanced version of the LS that improves LS with prior knowledge of the noise and second-order channel statistics. By minimizing the Euclidean distance between the channel gain at the pilot symbol, $\mathbf{H}$, and LS output, $\mathbf{H}_{LS}$, the LMMSE-based CE at pilot sub-carrier $p$ is given as \cite{LS+MMSE_literature}
    \begin{align}
        \mathbf{H}^{LMMSE}_{P} =\mathbf{R}_{H H_{P}}\times(\mathbf{R}_{H_{P} H_{P}} + {I}\times \frac{\sigma^2_{N}}{\sigma^2_{X}})^{-1} \times \mathbf{H^{LS}_{P}}
        \label{eqn:MMSE}
    \end{align}\normalsize
    where $\mathbf{R}_{H H_{p}}$ is the cross-correlation matrix between the channel at pilot symbol $\mathbf{H}$, and channel at pilot locations $\mathbf{H}_{p}$. ${\mathbf{R}_{H_p H_p}}$ is the auto-correlation matrix of $\mathbf{H}_{P}$, and $\frac{\sigma^2_{N}}{\sigma^2_{X}}$ represents the reciprocal of the signal-to-noise ratio (SNR). In the absence of prior knowledge, $\mathbf{R}_{H H_{p}}$ and ${\mathbf{R}_{H_p H_p}}$ can be obtained using past channel measurements. Though LMMSE offers superior performance over LS, the computational complexity and latency are high. Please refer to Section \ref{Sec: PA} for more details.

 \section{State-of-the-art DL-based CE}
\begin{figure}[!h]
    \centering
    \includegraphics[scale=0.50]{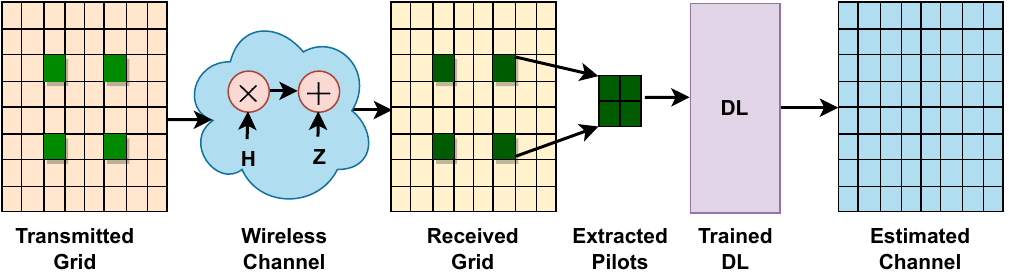}
    \caption{DL-based channel estimation.}
    \label{fig:Data_process}
\end{figure}

Recently, few works have explored advances in DL for CE in wireless PHY where statistical CE is completely replaced with the DL architecture, as shown in Fig.~\ref{fig:Data_process}. This section briefly reviews these DL approaches for the proposed work on the realization of existing DL-based CE on the SoC.


\subsubsection{ChannelNet}
\begin{figure}[!h]
    \centering
    \includegraphics[scale=0.7]{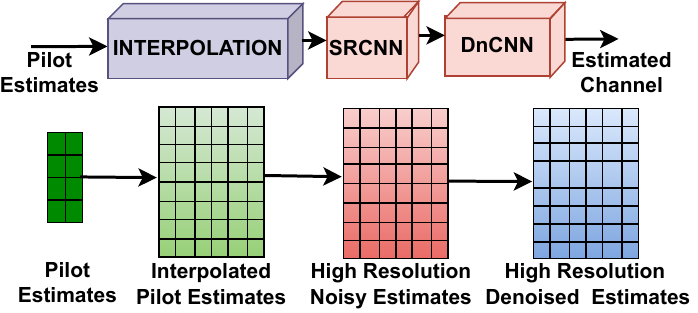}
    \caption{ChannelNet architecture in \cite{channelNet} for channel estimation.} 
    \label{fig:ChannelNet}
\end{figure}

ChannelNet \cite{channelNet} is one of the first DL-based CE approaches and as shown in Fig.~\ref{fig:ChannelNet}, it treats the CE as an image reconstruction problem. The channel estimates at the pilot sub-carriers are considered a low-resolution noisy input image, and the task is to reconstruct a high-resolution image that covers the entire OFDM data frame. First, bicubic interpolation is used to approximate channel matrix, followed by two DL architectures: Super-Resolution Convolutional Neural Network (SRCNN) \cite{SRCNN} and Denoising Convolutional Neural Networks (DnCNN) \cite{DNCNN}. The SRCNN comprises three convolutional layers to reconstruct the high-resolution image representing the noisy channel estimates at all sub-carriers. These estimates are passed through DnCNN, which consists of 20 convolutional layers, effectively reducing noise and improving the reliability of the channel estimates. Though ChannelNet offers superior CE performance than LS, it has high computational complexity. ChannelNet has  670,000 learnable parameters and 684 million multiplication and accumulation (MAC) operations for the PHY specifications considered in this paper.




\subsubsection{ReEsNet}
\begin{figure}[!h]
    \centering
    \includegraphics[scale=0.60]{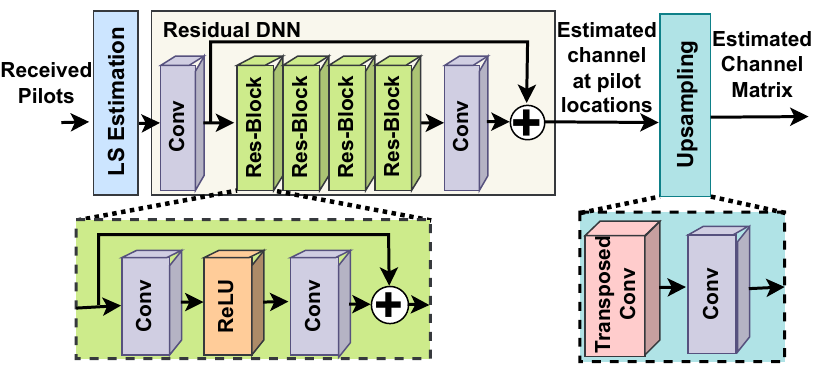}
    \caption{ReEsNet architecture in \cite{cite:ReEsNet} for channel estimation.}
    \label{fig:ReEsNET}
\end{figure}

In \cite{cite:ReEsNet}, residual neural network (ResNet) based CE, referred to as ReEsNet, is proposed where the interpolation/upsampling is moved after the DNN compared to ChannelNet. 
The LS output is fed as input to a single residual DNN based on the single image super-resolution network in \cite{DNNforReEsNet}. It comprises two convolutional layers with four residual blocks between them, as shown in Fig.~\ref{fig:ReEsNET}. Each residual block consists of two embedded convolutional layers with a rectified linear unit (ReLU) activation function in between. After the DNN, interpolation/upsampling is performed using a transposed convolutional layer followed by a convolutional layer. ReEsNet achieves improved performance compared to ChannelNet since it takes LS-based CE as input and use efficient single residual DNN before interpolation/upsampling. The ReEsNet needs around 53,000 learnable parameters and 32 million MAC operations, offering 10 times lower complexity than ChannelNet. 


\subsubsection{iResNet}\label{sec:iResNet}
\begin{figure}[!h]
     \centering
     \includegraphics[scale=0.55]{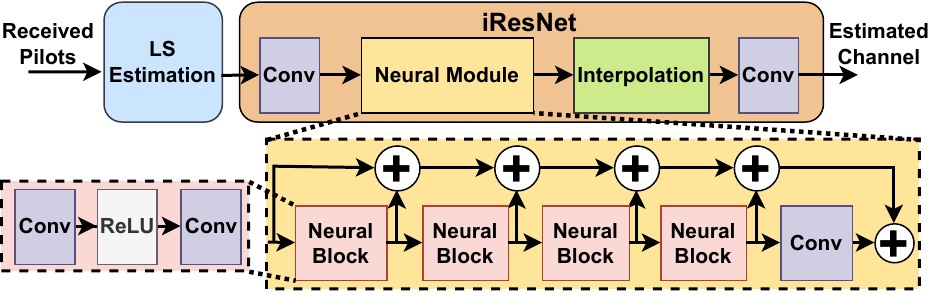}
     \caption{iResNet architecture in \cite{iResNet} for channel estimation.}
     \label{fig:Inter-ResNet strcucture}
 \end{figure}

The interpolation-ResNet \cite{iResNet} \cite{Channelformer} is an enhanced version of ReEsNet with lower complexity and nearly identical CE performance. As shown in Fig.~\ref{fig:Inter-ResNet strcucture}, iResNet optimizes the interpolation task in ResNet by replacing the computationally complex transposed convolution layer in interpolation with conventional bi-linear interpolation. In addition, residual blocks in ReEsNet are modified and renamed as neural blocks. Each neural block consists of two convolutional layers with a ReLU activation function in between, as in residual blocks. However, the skip connection between neural blocks is removed. Instead, the input and output of each neural block are accumulated and added to the input of the interpolation layer. For the PHY layer specifications discussed in this paper, the iResNet needs only 9442 learnable parameters and 4 million MAC operations, offering eight times lower complexity than ReEsNet with nearly identical performance.

\section{Proposed DL-Augmented Channel Estimation}
\label{Sec:LSiDNN}
The main challenge with iResNet is a huge number of MAC operations, which results in high complexity, latency, and power consumption. From the memory bandwidth perspective, iResNet involves a significant number of iterative read and write from memory due to convolution operations. As discussed later in Section~\ref{Sec: PA}, the latency of iResNet is significantly high, which may limit its usefulness for practical wireless PHY. We aim to address these drawbacks via a novel DNN-based CE approach.

\subsection{Proposed LSiDNN} \label{sec:LSDNN}
The proposed LSiDNN is based on a novel approach in which LS-based CE is augmented with the DNN. Compared to existing works that use separate DL for CE and interpolation, we use the DNN to improve the LS output and perform the 2-D interpolation simultaneously.

\begin{figure}[!h]
    \centering
    \includegraphics[scale=0.60]{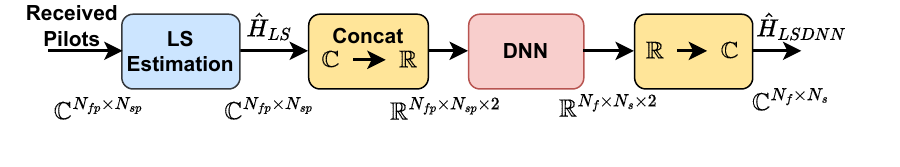}
    \caption{Proposed LSiDNN approach for channel estimation.}
    \label{fig:LSDNN_FLOW}
\end{figure}

As shown in Fig.~\ref{fig:LSDNN_FLOW}, the input to LSiDNN is the LS estimated channel at pilot locations denoted as $\hat{H}_{LS}\in \mathbb{C}^{N_{fp}\times N_{sp}} $ where $N_{sp}$ and $N_{fp}$ denote the number of OFDM symbols containing pilot and number of pilot sub-carriers in an OFDM symbol. After the LS-based CE, the 2D estimates are flattened to a 1D representation. Then, the real and imaginary components are concatenated, resulting in a single real-valued vector $\in \mathbb{R}^{\left ( N_{fp} \times N_{sp} \times 2 \right ) \times 1}$. The LSiDNN comprises of a three-layered, fully connected neural network. The input layer has a size of $N_{fp} \times N_{sp} \times 2$, representing the flattened and concatenated LS estimates. The hidden layer is equal to half of the input layer size, while the output layer has a size of $N_f \times N_{s} \times 2$, corresponding to the entire channel matrix. In the hidden layer, the ReLU activation function is employed to introduce non-linearity and enhance the learning capabilities of the network. However, the output layer does not utilize any activation function, as the purpose is to directly obtain the channel matrix estimates without further non-linear transformations. The output of LSiDNN is the flattened and concatenated channel matrix $\in \mathbb{R}^{\left ( N_f \times N_{s} \times 2 \right ) \times 1}$. These estimates are subsequently converted back into the complex domain and rearranged to obtain the final channel matrix denoted as $\hat{H}_{LSDNN} \in \mathbb{C}^{N_f\times N_s}$. 

All DL-based CE discussed in this paper are trained offline to minimize the error in channel estimates at the pilot locations and to interpolate the estimates in both the time and frequency dimensions to obtain the channel matrix for the complete data frame. We use the channel's impulse response as a reference for training, and in case impulse response is not available, we can explore a high SNR received signal as a reference. All experiments and results presented in this paper are done using Matlab with a dataset consisting of 10,000 OFDM frames for each type of wireless channel. This dataset is divided into training and validation sets, following an 80:20 ratio. Specifically, 8,000 OFDM frames are used for training, while the remaining 2,000 are allocated for validation. Additionally, a separate dataset comprising 1,000 OFDM frames is used for testing during inference. During the training phase, the DNN is trained for 250 epochs. A batch size of 256 and a learning rate of 0.01 are used in the training process. The ADAM optimizer is employed to optimize the network parameters and improve the training efficiency. Mean squared error (MSE) is used as a loss function.

\begin{figure}[!h]
    \centering
    \includegraphics[scale=0.415]{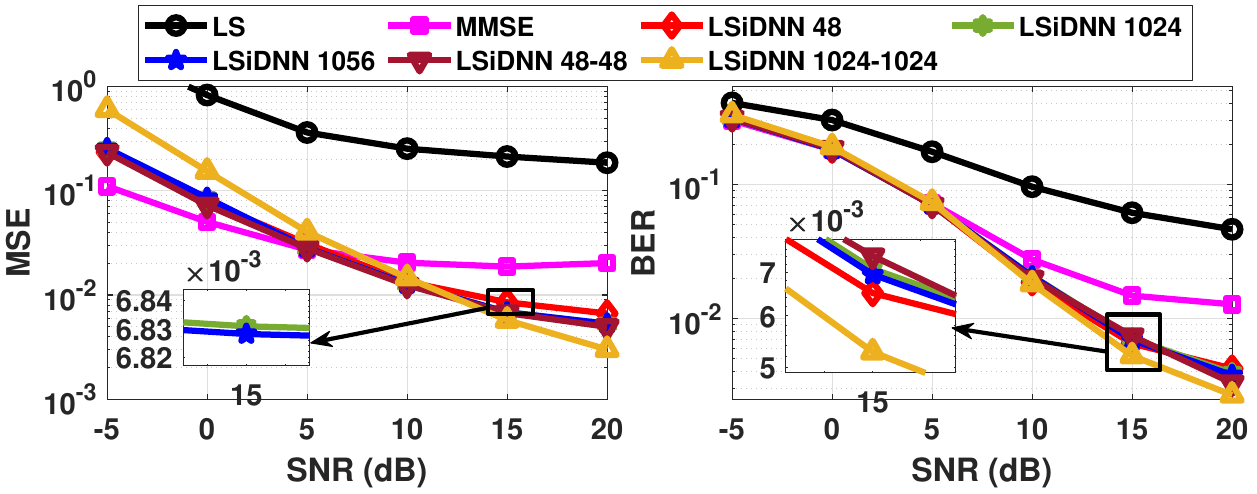}
    \caption{MSE and BER performance of different LSiDNN architectures.}
    \label{fig:ablation}
\end{figure}

In Fig.~\ref{fig:ablation}, we compare the BER performance of various configurations of the proposed LSiDNN. Specifically, we consider 5 configurations: 1) LSiDNN with a single hidden layer of 48 neurons (LSiDNN 48), 2) LSiDNN with a single hidden layer of 1024 neurons (LSiDNN 1024), 3) LSiDNN with a single hidden layer of 1056 neurons (LSiDNN 1056), 4) LSiDNN with two hidden layers of sizes 48-48 (LSiDNN 48-48), and 5)   LSiDNN with two hidden layers of sizes 1024-1024 (LSiDNN 1024-1024). The input and output sizes remain constant across all cases, as they depend on the pilot arrangement and frame size, respectively.
The performance of various LSiDNN configurations is nearly identical, and we can observe minor improvement in BER with the increase in DNN size. However, such minor improvement results in a significant penalty in the complexity as shown in Table~\ref{tab:ablation}. We have selected LSiDNN 48 architecture as it offers an optimal balance between performance and complexity.


\begin{table}[!htbp] 
\centering
\caption{\small Theoretical complexity analysis of different architectures for LSiDNN}
\begin{tabular}{|l|c|c|}
\hline
\textbf{Architecture} & \multicolumn{1}{l|}{\textbf{\# Learnable Parameters}} & \multicolumn{1}{l|}{\textbf{\# MAC operations}} \\ \hline
\textbf{LSiDNN 48} & \textbf{103440} & \textbf{101376} \\ \hline
\textbf{LSiDNN 1024} & 2165728 & 2162688 \\ \hline
\textbf{LSiDNN 1056} & 22333344 & 2230272 \\ \hline
\textbf{LSiDNN 48-48} & 105792 & 103680 \\ \hline
\textbf{LSiDNN 1024-1024} & \textbf{32153288} & \textbf{3211264} \\ \hline
\end{tabular}
\label{tab:ablation}
\end{table}

\subsection{Performance and Complexity Analysis for Software Realizations Using Floating Point Arithmetic} \label{sec:SimPerf}

In this section, we compare the performance of the proposed LSiDNN with conventional and DL-based approaches using double precision floating point (DPFL) wordlength (WL). We use mean squared error (MSE) and bit error rate (BER) as performance metrics. MSE indicates how closely the channel estimates align with the actual channel matrix, while BER measures the end-to-end performance of wireless PHY. Various specifications of wireless PHY are given in Table~\ref{tab:BASEBAND_PARAMETERS}. We consider a wide range of SNRs, different Doppler shifts, and 3GPP's multi-path fading channels: Extended Pedestrian A model (EPA), Extended Vehicular A model (EVA), and Extended Typical Urban model (ETU). These channels differ in their delay profiles \cite{Channel_models}. The EPA channel model represents a low delay spread environment, whereas the ETU model corresponds to a high delay spread environment, as shown in Table~\ref{tab:channel_model_delay_profile}. For each SNR value, corresponding MSE and BER results are obtained after averaging over 1000 data frames.

 \begin{table}[!h]
\centering
\caption{Specifications of Wireless PHY}
\resizebox{\columnwidth}{!}{
\begin{tabular}{@{}|c|c|@{}}
\hline
\textbf{Parameter}                & \textbf{Particuar} \\ \hline
Modulation Type                   & QPSK               \\ \hline
Guard interval type               & Cyclic Prefix (CP) \\ \hline
Noise model                       & AWGN               \\ \hline
Pilot   Subcarriers               & 24                 \\ \hline
Pilot Symbols                     & 2                  \\ \hline
Number of   deployed subcarriers  & 72                 \\ \hline
CP Length                         & 16                 \\ \hline
Bandwidth                         & 1.08 MHz           \\ \hline
Carrier   frequency               & 2.1 GHz            \\ \hline
Subcarrier   Spacing              & 15 KHz             \\ \hline
Number of   frame per slot        & 1                  \\ \hline
Number of OFDM   symbols per slot & 14                 \\ \hline
\end{tabular}
}
\label{tab:BASEBAND_PARAMETERS}
\end{table}

\begin{table}[!h]
    \renewcommand{\arraystretch}{1.1}
    \label{tab:floatResults}
    \caption{Delay Profile of Three Wireless Channel Models}
    \resizebox{\columnwidth}{!}{
    \begin{tabular}{|c|l|c|c|c|c|c|c|c|c|c|}
    \hline
    \multirow{2}{*}{ETU} & PathDelays (ns)  & 0    & 50   & 120  & 200  & 230  & 500   & 1600  & 2300  & 5000  \\ \cline{2-11} 
                              & AveragePathGains & -1.0 & -1.0 & -1.0 & 0.0  & 0.0  & 0.0   & -3.0  & -5.0  & -7.0  \\ \hline
    \multirow{2}{*}{EPA}          & PathDelays (ns)  & 0    & 30   & 70   & 90   & 110  & 190   & 410   & --    & --    \\ \cline{2-11} 
                              & AveragePathGains & 0    & -1.0 & -2.0 & -3.0 & -8.0 & -17.2 & -20.8 & --    & --    \\ \hline
    \multirow{2}{*}{EVA}          & PathDelays (ns)  & 0    & 30   & 150  & 310  & 370  & 710   & 1090  & 1730  & 2510  \\ \cline{2-11} 
                              & AveragePathGains & 0    & -1.5 & -1.4 & -3.6 & -0.6 & -9.1  & -7.0  & -12.0 & -16.9 \\ \hline
    \end{tabular}
    }

\label{tab:channel_model_delay_profile}
\end{table}

 \begin{figure*}[!h]
    \centering
    \includegraphics[scale=0.50]{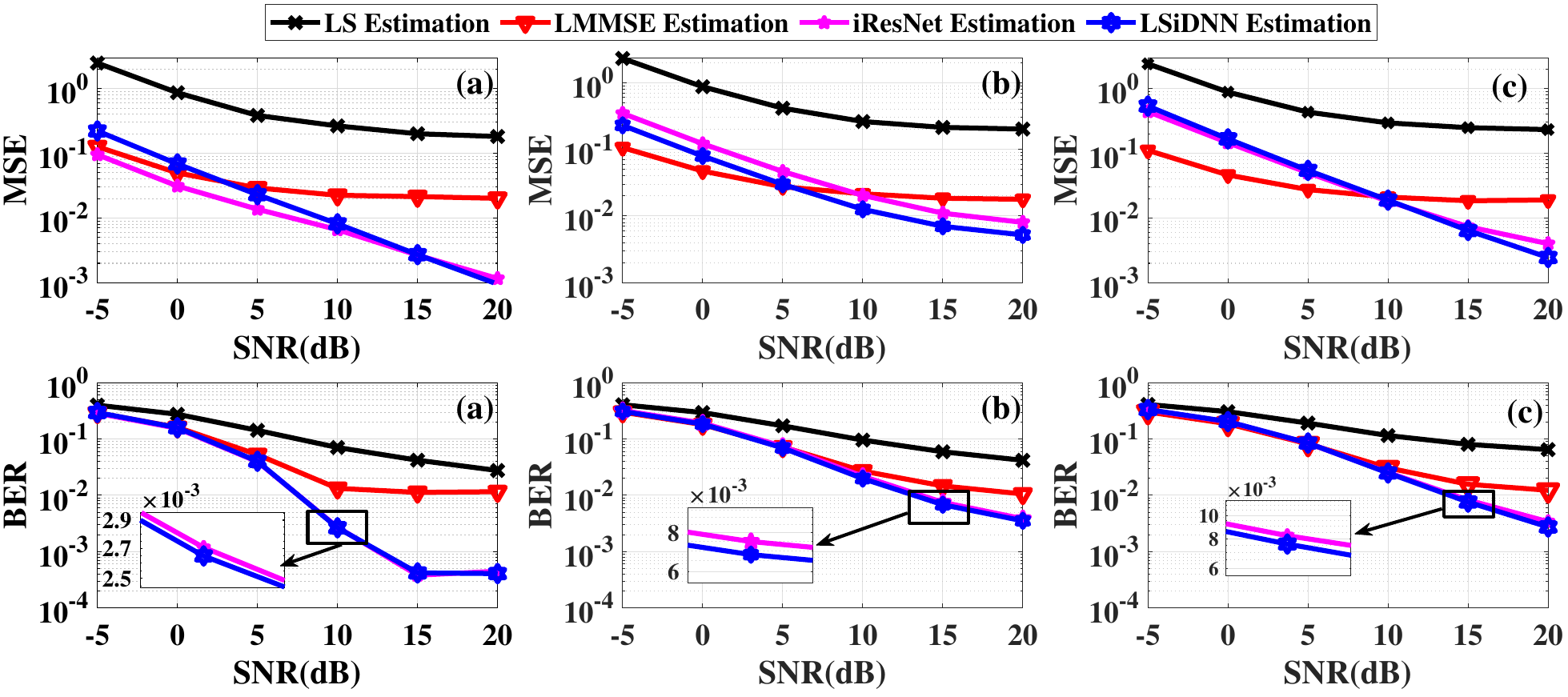}
    \caption{MSE and BER performance comparison of LSiDNN with other estimation techniques for (a) EPA channel, (b) EVA channel, and (c) ETU channel at maximum doppler shift of 97Hz. }
    \label{fig:perf_comp}
 \end{figure*}
\begin{figure*}[!h]
    \centering
    \includegraphics[scale=0.70]{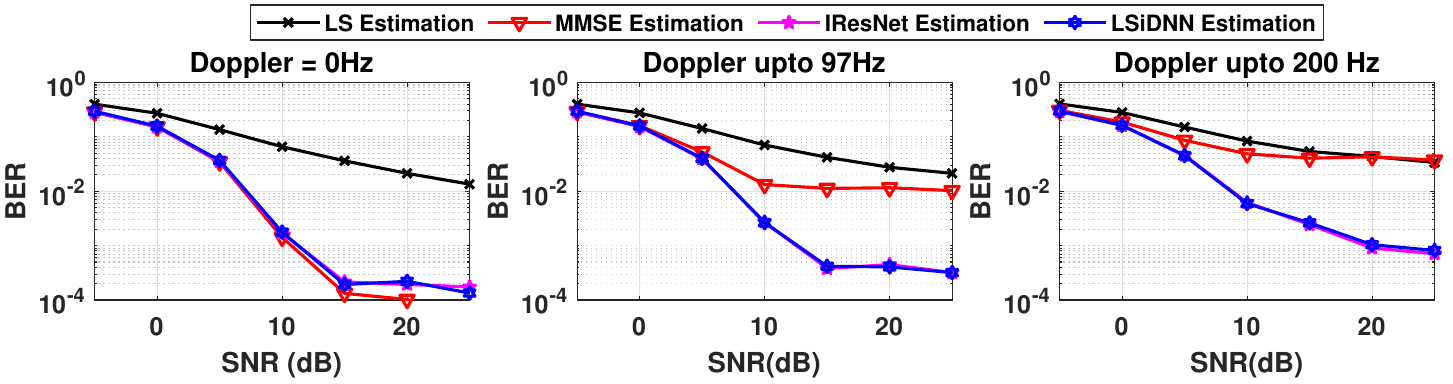}
    \caption{Effect of moblity on BER performance.}
    \label{fig:berDoppler}
 \end{figure*}
In Fig.~\ref{fig:perf_comp}, we compare the MSE and BER performance of various CE approaches for three different channels over the SNR range of -5 dB to 20 dB with a maximum velocity of 50kmph (i.e., Doppler shift of 97 Hz). As shown in Fig.~\ref{fig:perf_comp}, the performance of all CE approaches improves with increased SNR. The statistical LS and LMMSE approaches are significantly outperformed by DL-based CE approaches, especially at higher SNR. 
The performance of the iResNet and LSiDNN is nearly identical, with the proposed LSiDNN offering a slight improvement in BER, especially at high SNRs. For channels with low delay spread, such as EPA, the BER of LSiDNN and iResNet is almost identical. For channels with high delay spread, such as ETU, the BER of LSiDNN is lower than that of iResNet.

Next, we analyze the effect of Doppler velocity, i.e., mobility, on the performance of the CE. As Doppler velocity increases, the coherent time of the channel decreases. Consequently, the channel characteristics at the beginning of the frame may differ significantly from those at the end of the frame. This disparity leads to poor channel estimation performance with conventional approaches. The DL-based CE offers better performance, and the performance degradation due to increased Doppler velocity is less severe, as shown in Fig. \ref{fig:berDoppler}.

\subsection{Theoretical Complexity Analysis}\label{sec:compAnalysis}
In Table~\ref{tab:complexityComparison}, we compare the number of learnable parameters and MAC operations of various DL-based CE approaches. The number of learnable parameters impacts the model size and off-chip memory utilization, while the number of MAC operations determines the computational complexity and overall latency of the CE. Among existing state-of-the-art approaches, iResNet is the preferred approach, and it is considered as a benchmark along with LS and LMMSE in the rest of the paper. Proposed LSiDNN offers a 97\% reduction in MAC operations compared to iResNet. However, the number of learnable parameters in LSiDNN is more than 11 times that of iResNet, which impacts the off-chip memory. As discussed later in Section~\ref{sec:Comparison_Arch}, LSiDNN requires lower on-chip memory than iResNet because the convolution layers in iResNet require buffering and reuse of a large number of intermediate outputs. For any edge-computing platform, architecture with lower on-chip memory requirements is preferred since it is significantly more expensive and area-intensive than off-chip memory.

\begin{table}[!h]
\centering
\caption{\small Theoretical Complexity Comparison of DL-based CE}
\label{tab:complexityComparison}
\renewcommand{\arraystretch}{1.2}
 \resizebox{\columnwidth}{!}{
\begin{tabular}{|l|l|l|l|l|}
\hline
\textbf{}                        & \textbf{ChannelNet} \cite{channelNet} & \textbf{ReEsNet} \cite{cite:ReEsNet} & \textbf{iResNet} \cite{iResNet} & \textbf{LSiDNN} \\ \hline
\textbf{\# Learnable parameters} & 678K                & 53K              & \textbf{9K}      & 103K           \\ \hline
\textbf{\# MAC operations}       & 648M                & 32M              & 4M               & \textbf{0.1M}  \\ \hline
\end{tabular}}
\end{table}

\begin{figure}[!h]
    \centering
    \includegraphics[scale=0.4]{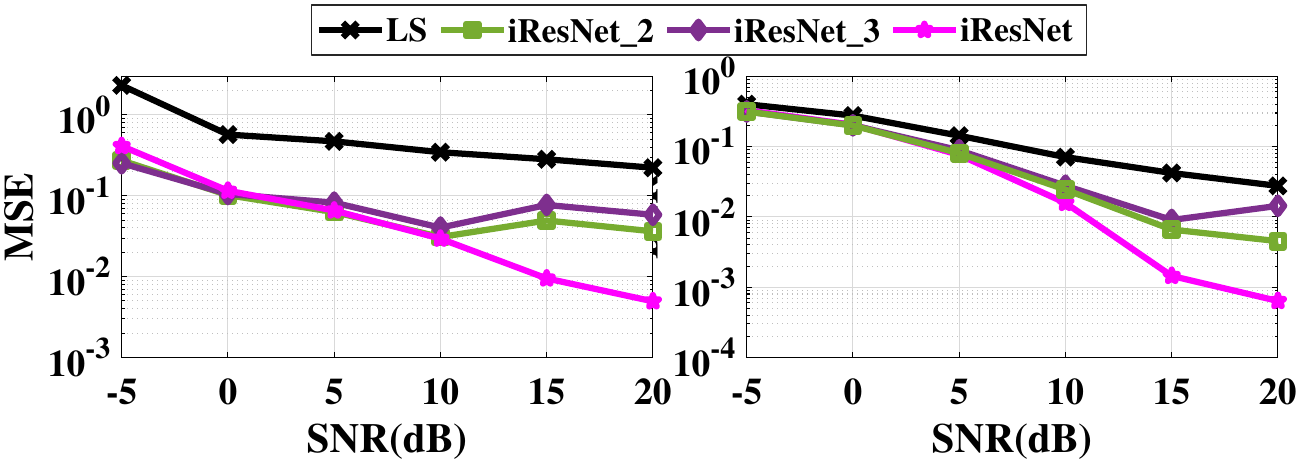}
    \caption{NMSE and BER performance of reduced complexity iResNet\cite{iResNet} channel estimation. }
    \label{fig:iResNetBlocks}
\end{figure}

To reduce the number of MAC operations in iResNet, we explored three architectures based on the number of neural blocks: iResNet\_2, iResNet\_3, and the original iResNet, having 2, 3, and 4 neural blocks, respectively. Each variation was trained individually to analyze its performance. As shown in Fig.~\ref{fig:iResNetBlocks}, reducing the number of neural blocks results in a significant degradation in iResNet's performance. This degradation is severe for wireless channels with high delay spread, such as ETU. This study suggests that further reduction in iResNet's complexity is not feasible.

\section{Algorithms to Architecture Mapping of CE on SoC} \label{sec:socArch}
\begin{figure}[!h]
    \centering
    \includegraphics[scale=0.45]{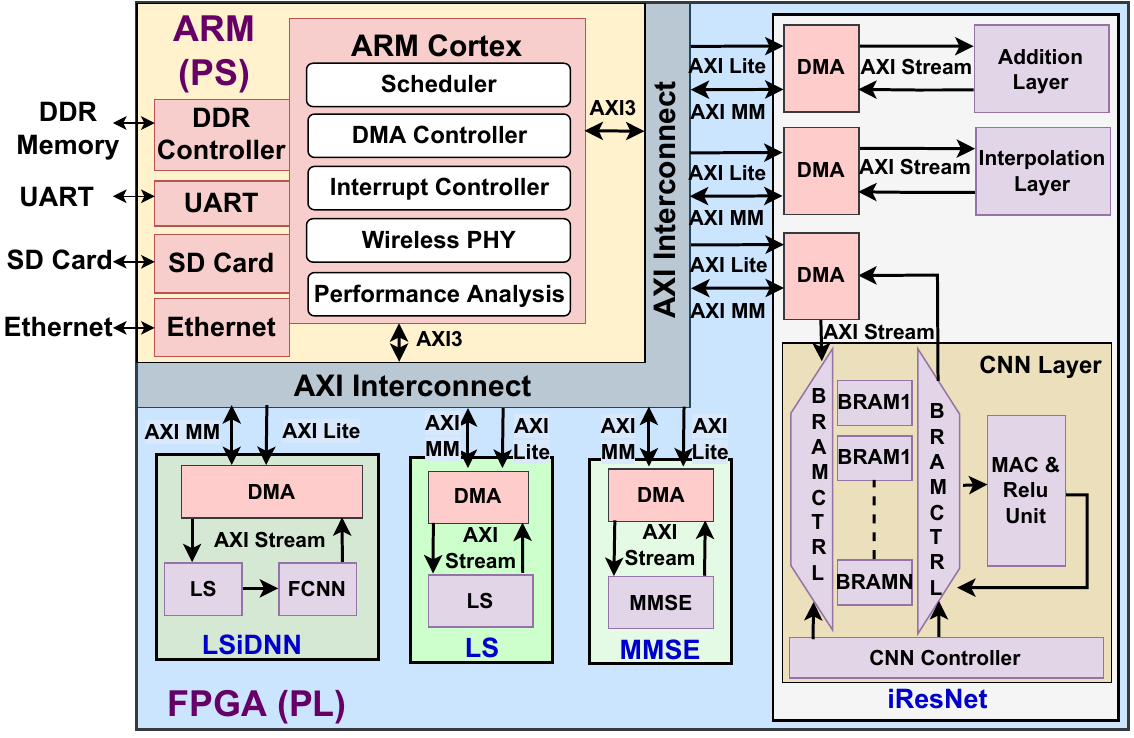}
    \caption{Hardware software co-design architecture of various CE approaches on ZSoC.}
    \label{fig:sysDesign}
\end{figure}

In this section, we propose an efficient mapping of existing and LSiDNN CE algorithms on SoC architectures. All hardware IPs are designed with Advanced eXtensible Interface (AXI), which makes them portable and easy to integrate in any SoC. We have used the Zynq SoC from AMD-Xilinx as a hardware platform consisting of a processing system (PS) and programmable logic (PL). The PS is based on an ARM processor, and the PL is FPGA. The communication between PS and PL is via the AXI interface, and we use direct memory access (DMA) for efficient data transfer between the PS cache, external DDR memory, and PL. The illustrative architecture with all CE in PL is shown in Fig.~\ref{fig:sysDesign}. The control and scheduling tasks are realized in PS. Since we focus on CE, the rest of the signal processing in wireless PHY is also realized in PS. In Section \ref{sec:HSCD}, we explore various hardware-software co-design configurations by dynamically allocating hardware blocks of CE between the PS and PL to study the effect on resource utilization, execution time, and power consumption. Next, we discuss the proposed hardware architectures of various CE approaches.

\subsection{Conventional Channel Estimation}


LS estimation is a computationally efficient technique for CE due to simple computations and low memory requirements. It is widely used in commercial applications, and various hardware realizations on SoC are available \cite{haq2022LSDNN}.

\begin{figure}[!b]
    \includegraphics[scale=0.55]{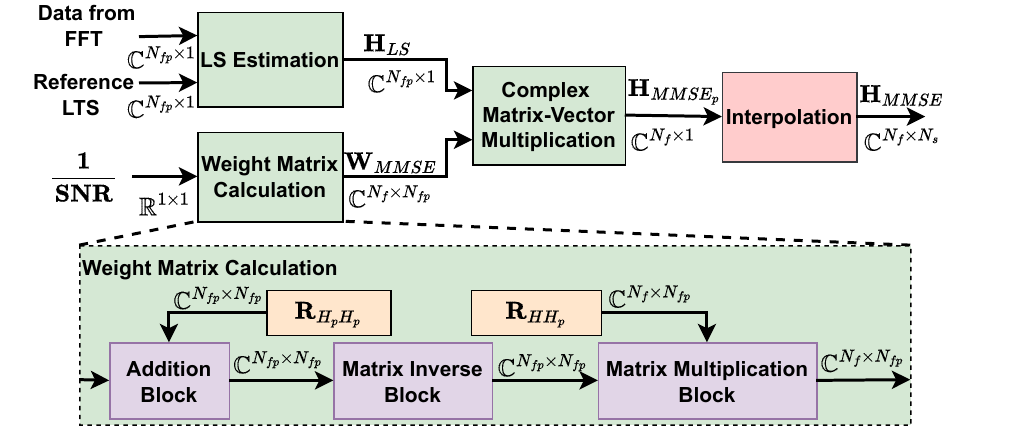}
    \caption{LMMSE hardware architecture}
    \label{fig:MMSE_HW_arch}
\end{figure}

LMMSE is another popular approach involving channel correlation matrices and SNR information, as discussed in Eq.~\ref{eqn:MMSE}. Since the channel is unknown, we can not calculate the channel matrix in real systems. Hence, correlation matrices are computed offline using prior channel matrices and stored in memory for real-time LMMSE estimation. This makes the LMMSE estimation highly sensitive to channel statistics. Also, it is computationally complex, and from Eq.~\ref{eqn:MMSE}, we can observe that the arithmetic operations in MMSE are not hardware-friendly. 

The architecture of the LMMSE involves the LS estimation followed by multiplication with weight matrix as shown in Fig.~\ref{fig:MMSE_HW_arch}. The calculation of the MMSE weight matrix, $\textbf{W}_{MMSE}$, involves the addition of the inverse of the SNR to the diagonal elements of the channel auto-correlation matrix $R_{H_pH_p}$. To optimize computational efficiency, we add $\frac{1}{SNR}$ only to the real part of the complex number. The output of this block is then fed to a matrix inverse module of size  \(N_{fp} \times N_{fp}\), where a complex matrix inversion operation is performed. The resulting inverse matrix is then multiplied with the channel auto-correlation matrix to obtain the LMMSE weight matrix. The weights matrix is subsequently multiplied with the LS estimated channel vector to obtain the 1D-MMSE estimate, $\mathbf{H}_{MMSE_p}$. Finally, time interpolation is performed to obtain estimates for the entire OFDM time-frequency grid.

The matrix inversion through the method of cofactors is unfeasible due to its excessively high computation time. In this work, we have explored three matrix inversion approaches: 1) Gauss-Jordan \cite{MMSE_GJ}, 2) QR\cite{MMSE_QR}, and 3) LU decomposition \cite{MMSE_LU}. The hardware architectures of these LMMSE algorithms are designed by adapting the reference matrix inversion examples provided by AMD-Xilinx for complex matrices. Since LU offers better performance and lower complexity, the discussion is limited to LU-based LMMSE.

\subsection{Proposed IReSNet Hardware Architecture} \label{sec:iresnet_arch}

In this work, we perform an algorithm to architecture mapping of the iResNet-based CE discussed in Section \ref{sec:iResNet}. It consists of mainly three hardware blocks: 1) Convolution layer (CONV), 2) Addition, and 3) Interpolation. As shown in Fig.~\ref{fig:Inter-ResNet strcucture}, these three hardware blocks are used multiple times during the single execution of iResNet to obtain CE output, and the corresponding scheduler is realized in PS. In this section, we focus on the convolution block, the most computationally and memory-intensive block in iResNet. 

\begin{figure}[!h]
    \includegraphics[scale=0.375]{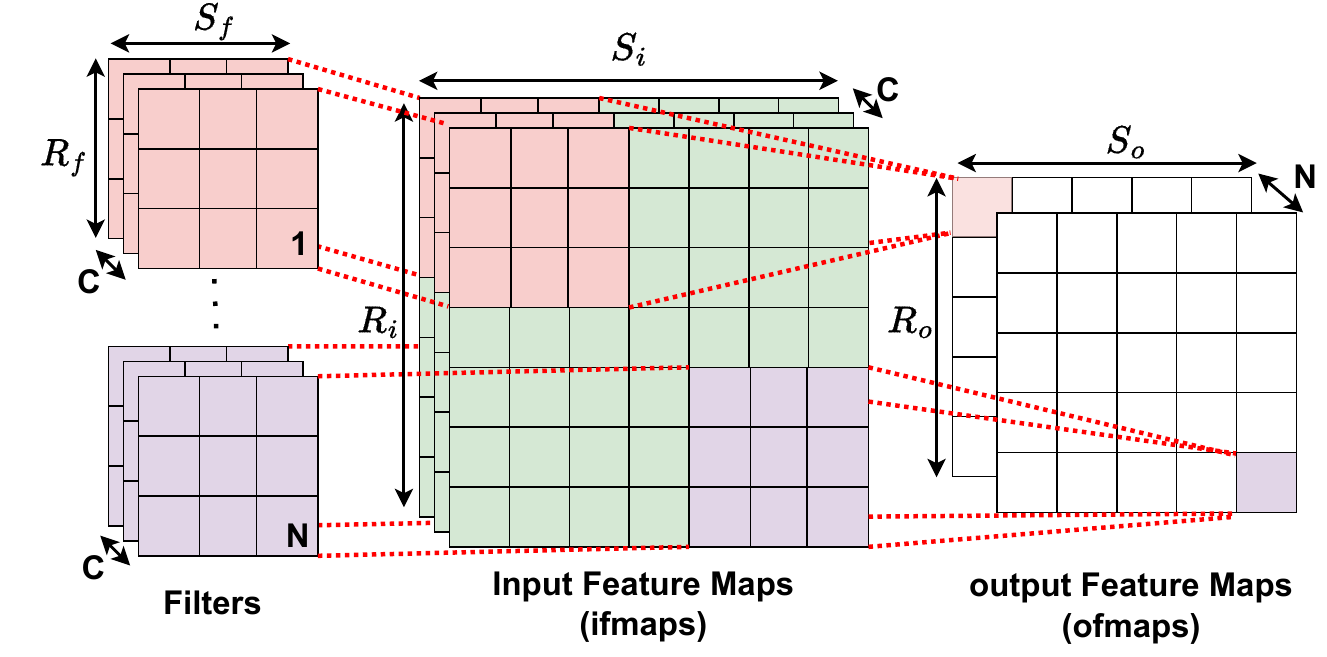}
    \caption{Convolution operation in conv layer}
    \label{fig:convOperation}
\end{figure}

CONV layer involves a convolution operation between a 3D filter kernel and the receptive field on the 3D input feature map (ifmap), as shown in Fig.~\ref{fig:convOperation}. The convolution operation involves element-wise multiplication followed by accumulation to generate an output feature map (ofmap). This convolution operation is performed over the entire ifmap by sliding the filter kernel in vertical and horizontal directions with a certain fixed stride to obtain a 2D ofmap. Multiple filters are employed to process each layer, and the convolution of each filter with the ifmap produces one channel of the ofmap. Hence, the convolution of a 4D kernel (multiple 3D filters) and a 3D input feature map generates a 3D output feature map. Mathematically,

\begin{equation} \label{eqn:CNN}
\begin{split}
 O[n][x][y] = B[n] +  \sum_{k=0}^{C-1} \sum_{i=0}^{R_f-1} \sum_{j=0}^{S_f-1} &I[k][Ux+i][Uy+j] \\
 & \times W[n][k][i][j]
\end{split}
\end{equation}
where $n$, $x$, and $y$ represent the output channel, row index, and column index, respectively. Similarly, $k$, $i$, and $j$ correspond to the input channel, row index, and column index, respectively. The variables $O$, $I$, $W$, and $B$ represent the values of the ofmap, ifmap, weight, and bias, respectively. The variable $U$ represents the stride value.

\begin{figure*}[!t]
    \centering
    \includegraphics[scale=0.675]{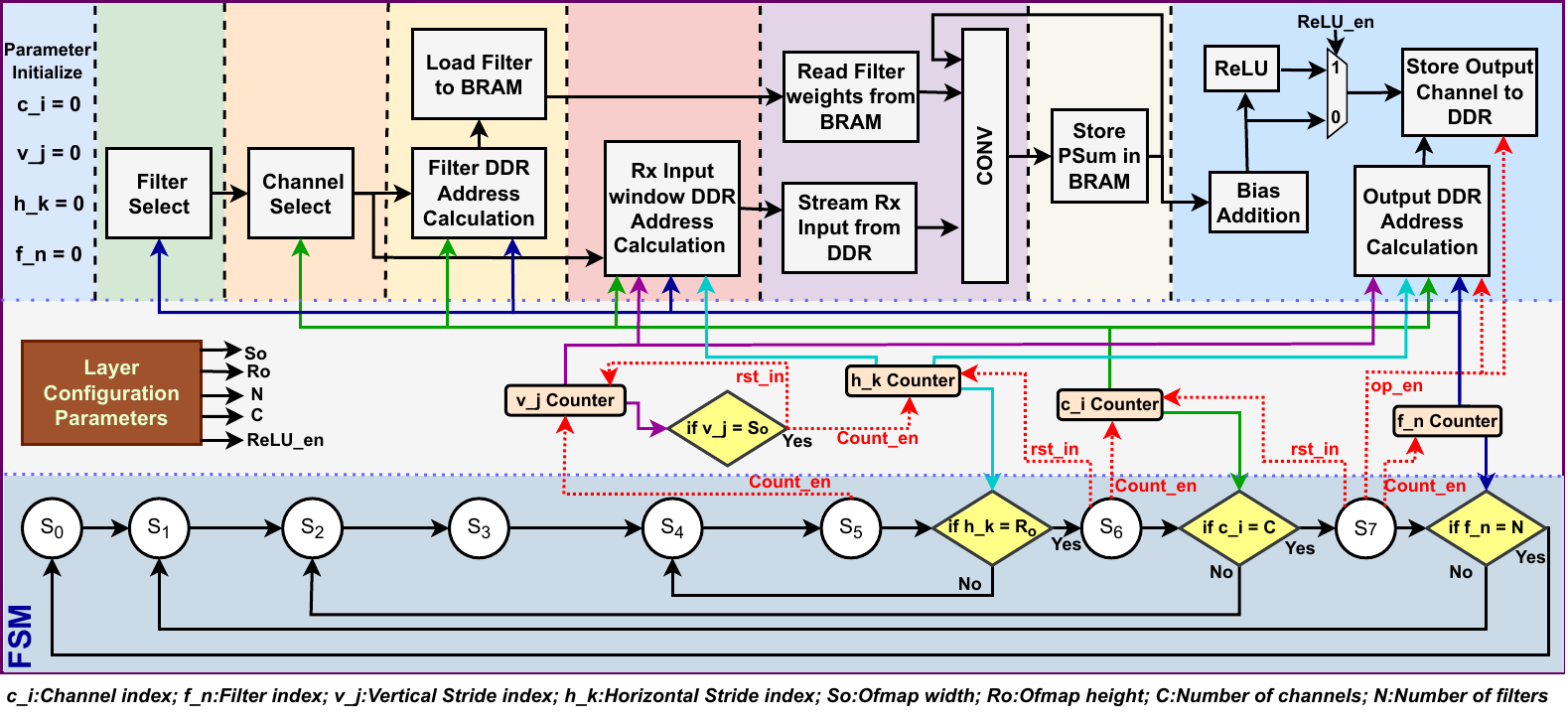}
    \caption{Proposed CONV layer architecture and schedular in the iReSNet.}
    \label{fig:iResNet_arch}
\end{figure*}


CONV layers are computationally intensive due to a large number of multiply-and-accumulate (MAC) operations. Additionally, they consume significant memory resources due to the generation of numerous intermediate outputs. The nature of convolution operations allows for potential acceleration, as MAC operations can be parallelized. However, memory access remains a bottleneck. CNN accelerators aim to reduce DDR access by leveraging the data reuse properties of CONV layers \cite{Sze2017DLsurvey}. The filter weights and bias values used for various convolution layers of multiple instances are stored in external DDR memory and accessed via DMA. We can not use on-chip memory on FPGA such as Block RAM since it is limited in size and convolution operations need temporary storage to buffer intermediate outputs, for which block RAM is the preferred choice.



The proposed architecture of the CONV layer is Fig.~\ref{fig:iResNet_arch} and it is based on the output stationary (OS) dataflow approach \cite{Reference_OutputStationary,OS_dataflow}.
In the OS dataflow, each processing element (PE) calculates a single output sample. The partial sums (psums) generated by MAC operations are stored in the register file (RF) of the same PE where they are produced, meaning the psums remain stationary within a PE. This approach reduces the cost of accessing psums. Weights and input pixels are streamed onto the PE without being stored. This results in fewer number of data movement of intermediate outputs.



As shown in Fig.~\ref{fig:iResNet_arch}, depending on the selected filter index, $f_n$ and channel index, $c_i$, the filter weights are read from external DDR memory and stored in BRAM in FPGA. These filters are repeatedly used for convolution operation with different windows of the input samples stream from DDR memory. This is accomplished using vertical and horizontal stride counters of appropriate size. In the end, appropriate bias is added. Depending on the use of CONV layer in iResNet, RELU is enabled, followed by writing the output of the CONV in the DDR. This process continues until all input channels have been processed and the final ofmap is computed.



The maximum number of  PEs in CONV layer of iResNet based CE is $N_f \times N_s$. However, different CONV blocks need different numbers of PEs and in the proposed architecture shown in Fig.~\ref{fig:sysDesign}, we have designed a single CONV block for various CONV operations. For CONV with fewer number of PEs, we have configured the proposed CONV IP on the fly using the AXI-Lite interface. To improve the performance, multiple instances of CONV IP can be instantiated in the FPGA depending on the desired trade-off between resource utilization and latency. Within each PE, the MAC operation is pipelined to reduce the latency without increasing the resource utilization on FPGA. It is possible to parallelize the multiplication operations in PE to reduce the execution time. However, it demands significant resource utilization and additional bandwidth to read the data from DDR in parallel for simultaneous data processing. Hence, we have chosen pipelined MAC operation in PE for this work.

\subsection{Proposed LSiDNN Architecture} \label{sec:lsidnn_arch}

\begin{figure}[!h]
    \centering
    \includegraphics[scale=0.7]{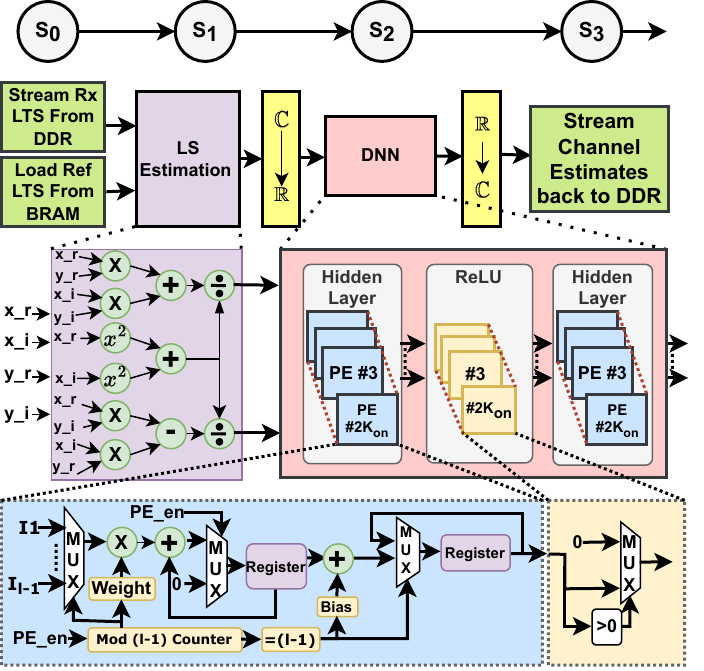}
    \caption{Proposed architecture for LSiDNN.}
    \label{fig:lsidnn_arch}
\end{figure}

The LSiDNN architecture consists of a cascade connection of the LS estimation module followed by an FCNN module, as discussed in Section \ref{sec:LSDNN}. In the LS estimation module, a complex division operation is performed between the received LTS and the reference LTS. This operation can be efficiently implemented in hardware using six real multiplication operations, two division operations, and three addition operations \cite{haq2022LSDNN}, as illustrated in the Fig.~\ref{fig:lsidnn_arch}. Additional optimization considers that the reference LTS is a BPSK modulated signal, simplifying the estimation process to selecting either the received complex value or its two’s complement based on whether the LTS is +1 or -1. Subsequently, the LS estimation module's output is flattened, and the real and imaginary components are concatenated before being fed into the FCNN.

The FCNN is structured in layers, each containing multiple neurons implemented as PEs. These PEs perform MAC operation between the outputs of the previous layer and their corresponding weight vectors, followed by bias addition and ReLU activation. Each PE in a layer performs a total of $P_{l-1}$ MAC operations, with $P_l$ representing the number of neurons in the $l^{th}$ layer. This requires memory to store the corresponding weight vector, a register to hold one bias value, and another register to store intermediate output. While a fully parallel MAC operation implementation is possible, it would demand significant computational resources and increased power consumption. Therefore, operations within a PE are serialized and pipelined. The implementation of the ReLU activation function is straightforward and involves selecting the maximum value between zero and the PE output. This selection can be achieved using a multiplexer with zero and the PE output as inputs, with the sign bit of the PE output serving as the select line as shown in Fig.~\ref{fig:lsidnn_arch}.


We have considered two versions of the LSiDNN architecture depending on the level of parallelization. In the low-latency architecture (LSiDNN-LL), we have adopted layer-level parallelization by implementing all PEs within a layer in parallel. The output from the previous layer is serially broadcasted to all PEs within the next layer, which reduces internal bandwidth requirements and allows all PEs to initiate operations simultaneously. The memory used to store outputs is partitioned to facilitate simultaneous access by all PEs within a layer. For resource-constrained devices, a computationally efficient architecture (LSiDNN-CE) is proposed, where a single PE performs the implementation of all neurons within a layer in a serial manner. As the weight vector dimensions for all neurons within a layer are the same, a fixed architecture can implement all neurons within that layer. In this case, the weight memory of such a PE requires $P_{l-1} \times P_l$ memory elements to store the entire weight matrix for each neuron. Neuron selection reduces to the selection of the appropriate weight vector. This architecture conserves computational resources at the expense of increased latency. Section \ref{Sec: PA} discusses the latency and hardware resource utilization of these two architectures.

As shown in Fig.~\ref{fig:sysDesign}, we have developed an AXI Stream-based hardware IP for LSiDNN. Given the low-complexity nature of LSiDNN's FCNN, we can store all weights and bias parameters in the internal BRAM memory, allowing for the concurrent implementation of all layers on the PL. This eliminates the need for scheduling layer operations and transferring weights from external memory, resulting in reduced power and complexity. 

\section{Performance and Complexity Analysis on SoC}
\label{Sec: PA}

 In this section, we present the performance and complexity analysis of the four CE techniques, LS, LMMSE, iReSNet, and LSiDNN, discussed in Section \ref{sec:socArch}. As a hardware platform, we use Xilinx ZC706, a state-of-the-art Zynq series heterogeneous SoC. It consists of a dual-core ARM Cortex A9 processor as processing system (PS) and a 7-series FPGA with 1090 units of 18kB Block RAMs, 80 DSP48E units, 218600 units of 6-input look-up tables (LUTs), and 437200 flip-flops (FFs) as programmable logic (PL). We consider the effect of hardware-software co-design (HSCD) and fixed-point WL on the performance of the CE. We also explore serial-parallel architectures to study resource utilization and latency trade-offs. To avoid repetition of discussions, we do not discuss the results on DPFL and SPFL architectures separately since the software results are already discussed in Section~\ref{sec:LSDNN}, and there is a perfect match between hardware and software results for floating-point WL.


\subsection{Hardware Software Co-design (HSCD)}\label{sec:HSCD}

In HSCD, we design various configurations of the CE architecture by dividing them into PS and PL of the Zynq SoC. The received data from the analog-to-digital converter (ADC) can be buffered in external DDR or Block RAM on PL before being processed by receiver PHY. For performance analysis, we consider DDR memory to account for worst-case memory communication time. Since the majority of the receiver PHY blocks before and after CE are realized in PL, realizing CE in PS may lead to substantial data communication overhead between PS and PL. The PL can accelerate the arithmetic and logical operations by parallel processing, while sequential tasks can be efficiently realized on PS. The signal processing in PS helps to reduce the PL size and, hence, lower cost and power consumption. However, algorithms in PS may have higher execution time. Hence, careful analysis of HSCD for each architecture is crucial to determine the optimal configuration for given resource utilization and execution time trade-off. 


\subsubsection{LS based CE} \label{sec:HSCD_LS}
The LS-based CE involves LS estimation at pilot locations, i.e., 24*2 = 48 locations in the data frame, followed by the interpolation (INTP) to obtain the channel estimation at 72*12 = 864 locations. As shown in Table~\ref{tab:LS_HSCD}, we designed and implemented four configurations on Zynq SoC. In Rows 1 and 4, LS and INTP are entirely realized in PS and PL, respectively. In Rows 2 and 3, one is realized in PS, and the other is in PL. Note that realization in PS does not include the overhead between PS and PL since CE input is assumed in DDR instead of PL. It can be concluded that LS-based CE can be efficiently realized in PL since realization in PS may result in higher execution time.


        \begin{table*}[!h]
          \caption{HSCD for LS on Zynq SoC}
       \resizebox{\textwidth}{!}{ \begin{tabular}{|c|c|c|c|c|c|c|c|c|c|}
        \hline
        \textbf{S. No.} & \textbf{PS} & \textbf{PL} & \textbf{Execution Time (s)} & \textbf{BRAM} & \textbf{DSP} & \textbf{FF} & \textbf{LUT}& \textbf{Total Power (W)} & \textbf{Dynamic Power (W)} \\ \hline
        1 & LS + INTP & NA & 0.070028 & NA & NA & NA & NA & 1.57 & NA \\ \hline
        2 & LS & INTP & 0.098543 & 0 & 8 & 2045 & 3016 & 2.029 & 1.81 \\ \hline
        3 & INTP & LS & 0.071297 & 0 & 10 & 8801 & 3703 & 1.965 & 1.747 \\ \hline
        4 & NA & LS + INTP & 0.070528 & 0 & 12 & 10524 & 5605 & 2.209 & 1.987 \\ \hline
        \end{tabular}}
        \label{tab:LS_HSCD}
        \end{table*}

    \begin{table*}[h]
    \caption{HSCD for iReSNet\cite{iResNet} on Zynq SoC}
    \resizebox{\textwidth}{!}{\begin{tabular}{|c|c|c|c|c|c|c|c|c|c|}
    \hline
    \textbf{S. No.} & \textbf{PS}   & \textbf{PL}   & \textbf{\begin{tabular}[c]{@{}c@{}}Execution \\ Time (s)\end{tabular}} & \textbf{BRAM} & \textbf{DSP} & \textbf{FF} & \textbf{LUT} & \textbf{\begin{tabular}[c]{@{}c@{}}Total \\ Power (W)\end{tabular}} & \textbf{\begin{tabular}[c]{@{}c@{}}Dynamic \\ Power   (W)\end{tabular}} \\ \hline
    1 & {CONV+ADD+INTP} & - & 40.25 & - & - & - & - & - & - \\ \hline
    2 & {CONV+ADD} & {INTP} & 40.45 & 0 & 8 & 2045 & 3016 & 2.027 & 1.808 \\ \hline
    3 & {CONV+INTP} & {ADD} & 36.86 & 0 & 10 & 1024 & 1153 & 2.255 & 2.03 \\ \hline
    4 & {ADD+INTP} & {CONV} & 18.26 & 112 & 100 & 10042 & 9590 & {2.379} & {2.153} \\ \hline
    5 & {INTP} & {CONV+ADD} & 18.99 & 112 & 116 & 12009 & 12417 & {2.714} & {2.479} \\ \hline
    6 & {ADD} & {CONV+INTP} & 19.24 & 112 & 108 & 12352 & 12648 & {2.486}
 & {2.258} \\ \hline
    7 & - & {CONV+ADD+INTP} & 19.26 & 112 & 126 & 14620 & 15244 & 2.852 & 2.615 \\ \hline
    \end{tabular}}
    \label{tab:IReSNet_HSCD}
    \end{table*}

    \begin{table*}[!h]
    \centering
    \caption{HSCD for LSDNN on Zynq SoC}
    \resizebox{\textwidth}{!}{\begin{tabular}{@{}|c|c|c|c|c|c|c|c|c|c|@{}}
    \hline
    \textbf{S. No.} & \textbf{PS}   & \textbf{PL}   & \textbf{\begin{tabular}[c]{@{}c@{}}Execution \\ Time (s)\end{tabular}} & \textbf{BRAM} & \textbf{DSP} & \textbf{FF} & \textbf{LUT} & \textbf{\begin{tabular}[c]{@{}c@{}}Total \\ Power (W)\end{tabular}} & \textbf{\begin{tabular}[c]{@{}c@{}}Dynamic \\ Power   (W)\end{tabular}} \\ \hline
    1 & {Layer 1+Layer 2} & - & 0.76345 & - & - & - & - & - & - \\ \hline
    2 & {Layer 2} & {Layer 1} & 0.723881 & 32 & 160 & 19638 & 15051 & 2.597	 & 2.34\\ \hline
    3 & {Layer 1} & {Layer 2} & 0.114685 & 264 & 32 & 12735 & 9773 & 2.612	 & 2.368\\ \hline
    4 & - & {Layer 1+Layer 2} & 0.08123 & 290 & 40 & 14735 & 10766 & 2.617 & 2.387\\ \hline
    \end{tabular}}
    \label{tab:LSDNN_HSCD}
    \end{table*}

\subsubsection{LMMSE based CE}
Since LMMSE is computationally complex among all CE approaches and involves huge data communication overhead with memory, we have not explored HSCD for LMMSE. 


\subsubsection{iReSNet}\label{HSCD_IReSNet}
The iReSNet architecture comprises three computational blocks: 1) the Convolution layer (Conv), 2) the Addition layer (ADD), and 3) the Interpolation layer (intp). We have explored 7 HSCD configurations for iReSNet, and corresponding results are shown in Table~\ref{tab:IReSNet_HSCD}. In Row 1, iReSNet is realized completely in PS with a total execution time of 40 $\mu$s. Moving interpolation and addition layers to PL does not offer a significant reduction in execution time, as shown in Rows 2 and 3. On the other hand, moving the data-intensive convolution layer (B1) offers more than two times reduction in execution times, as shown in Row 4. This comes with the additional resource utilization and power consumption of PL. As shown in Rows 5-7, moving B2 and B3 to FPGA along with B1 leads to a degradation in performance due to additional data communication overhead between PS and PL. Also, there is a further increase in resource utilization and power consumption. Thus, iReSNet architecture in Row 4 with only B1 in PL is preferred.



\subsubsection{LSiDNN}

The LSiDNN architecture consists of LS-based CE at pilot positions and two-layer DNN comprising L1 and L2 layers. Based on the results discussed in Section~\ref{sec:HSCD_LS}, LS is realized in PL. For DNN, we have explored four HSCD configurations as shown in Table~\ref{tab:LSDNN_HSCD}. It can be observed that the complete realization of both layers of DNN in PL (Row 4) offers $9\times$ improvement in execution time when compared to DNN in PS (Row 1). This considerable acceleration can be attributed to the abundant parallelization opportunities inherent in the computation of fully connected layers within the DNN architecture. The architecture with L2 in PL (Row 3) uses lower resources than Row 4 with slight degradation in execution time. For the rest of the discussion, Row 4 architecture is preferred.


\subsection{Fixed Point Word Length}
In Section \ref{sec:SimPerf}, we discussed the MSE and BER performance of the CE schemes using DPFL WL. Though the CE with DPFL WL is the most accurate due to the large dynamic range, the resource utilization, execution time, and power consumption are also high. The wireless transceiver PHY is usually deployed on edge devices with limited computational capability. Furthermore, it receives and sends the signals to data converters, which have a limited number of bits. Thus, the large dynamic range offered by DPFL may not be needed, so fixed-point architectures should be explored.


The PS does not support fixed-point WL; hence, only floating or integer WL representation is possible in PS. Since integer WL is unsuitable for CE, we can only have DPFL and SPFL WL in PS. On the other hand, PL, i.e., FPGA, can support any arbitrary WL. For illustration, we represent fixed-point WL as $(W, I)$ where $W$ is the total number of bits and $I$ is the number of bits out of $W$ to represent the integer part of the real number. Thus, $(W-I)$ is the number of bits representing the fractional part. To identify the minimum value of $W$, we first find the minimum $I$ for sufficiently large $W$ for a given dataset. Then, for the selected $I$, we find the minimum possible $(W-I)$ and hence, $W$. Since LMMSE performance degrades significantly even for SPFL compared to DPFL, we do not discuss the fixed-point architecture for LMMSE.


\subsubsection{Selection of $I$}\label{sec:I_sel}
To select minimum $I$, we compare the MSE of the fixed-point CE with that of SPFL. As shown in Table~\ref{tab:integer_bits}, the value of $I$ for LS is at least four since there is no improvement in MSE thereafter, and MSE with $I=4$ is the same as that of SPFL WL. Similarly, the minimum value of $I$ for iReSNet and LDiDNN is 4 and 8, respectively.

\begin{table*}[!h]
\caption{\small Selection of $I$ for LS, iResNet\cite{iResNet}, and LSiDNN.}
\renewcommand{\arraystretch}{1.2}
\label{tab:integer_bits}
 \resizebox{\textwidth}{!}{
 \begin{tabular}{|lll|lll|lll|}
\hline
\multicolumn{3}{|c|}{\textbf{LS}}                                                                                                    & \multicolumn{3}{c|}{\textbf{iResNet\cite{iResNet}}}                                                                                               & \multicolumn{3}{c|}{\textbf{LSiDNN}}                                                                                                \\ \hline
\multicolumn{1}{|l|}{\textbf{Word Length   (W,I)}} & \multicolumn{1}{l|}{\textbf{W-I}} & \textbf{Average MSE} & \multicolumn{1}{l|}{\textbf{Word Length   (W,I)}} & \multicolumn{1}{l|}{\textbf{W-I}} & \textbf{Average MSE} & \multicolumn{1}{l|}{\textbf{Word Length   (W,I)}} & \multicolumn{1}{l|}{\textbf{W-I}} & \textbf{Average MSE} \\ \hline
\multicolumn{1}{|l|}{\textbf{SPFL}}                                       & \multicolumn{1}{l|}{-}            & 0.29778              & \multicolumn{1}{l|}{\textbf{SPFL}}                                       & \multicolumn{1}{l|}{-}            & 0.099368             & \multicolumn{1}{l|}{\textbf{SPFL}}                                       & \multicolumn{1}{l|}{-}            & 0.125617             \\ \hline
\multicolumn{1}{|l|}{\textbf{(15,3)}}              & \multicolumn{1}{l|}{12}           & 0.52125              & \multicolumn{1}{l|}{\textbf{(15,3)}}              & \multicolumn{1}{l|}{12}           & 0.228793             & \multicolumn{1}{l|}{\textbf{(25,7)}}              & \multicolumn{1}{l|}{18}           & 0.249926             \\ \hline
\multicolumn{1}{|l|}{\textbf{(16,4)}}              & \multicolumn{1}{l|}{12}           & 0.29778              & \multicolumn{1}{l|}{\textbf{(16,4)}}              & \multicolumn{1}{l|}{12}           & 0.099639             & \multicolumn{1}{l|}{\textbf{(26,8)}}              & \multicolumn{1}{l|}{18}           & 0.125617             \\ \hline
\multicolumn{1}{|l|}{\textbf{(17,5)}}              & \multicolumn{1}{l|}{12}           & 0.29778              & \multicolumn{1}{l|}{\textbf{(17,5)}}              & \multicolumn{1}{l|}{12}           & 0.099891             & \multicolumn{1}{l|}{\textbf{(27,9)}}              & \multicolumn{1}{l|}{18}           & 0.125617             \\ \hline
\end{tabular}
 }
 \end{table*}

\subsubsection{Selection of $W$}\label{sec:W_Sel}
\begin{figure}[!h]
    \centering
    \includegraphics[scale=0.34]{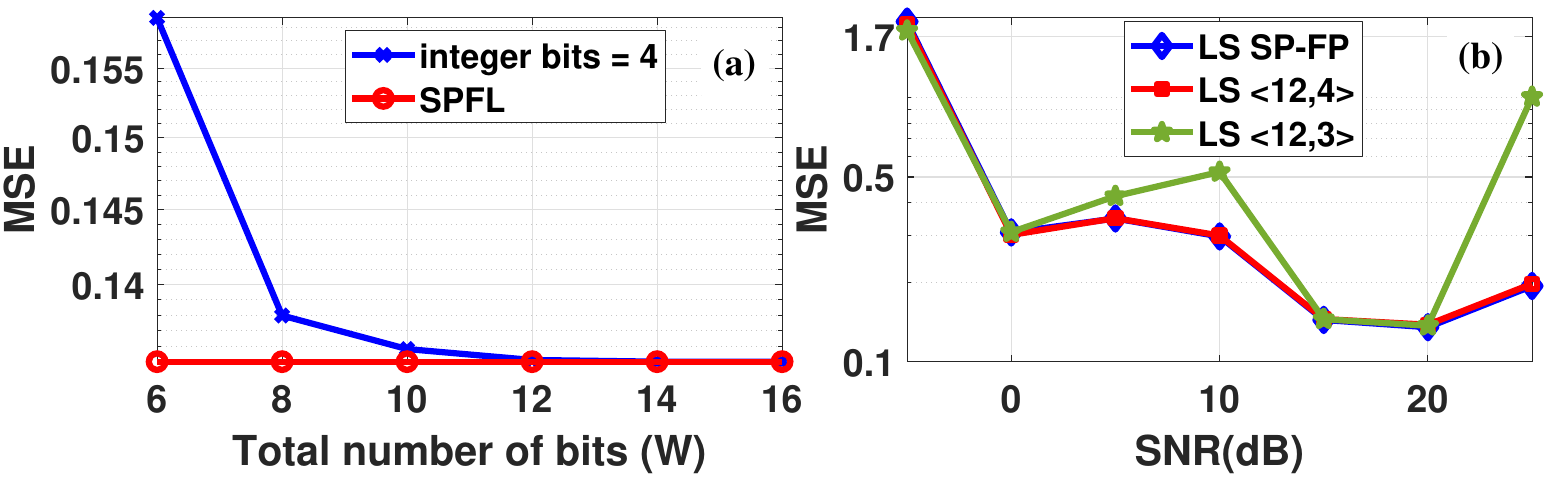}
    \caption{Effect of word length on the MSE performance of LS for fixed $I=4$.}
    \label{fig:LS_WL}
\end{figure}

\begin{figure}[!h]
    \centering
    \includegraphics[scale=0.33]{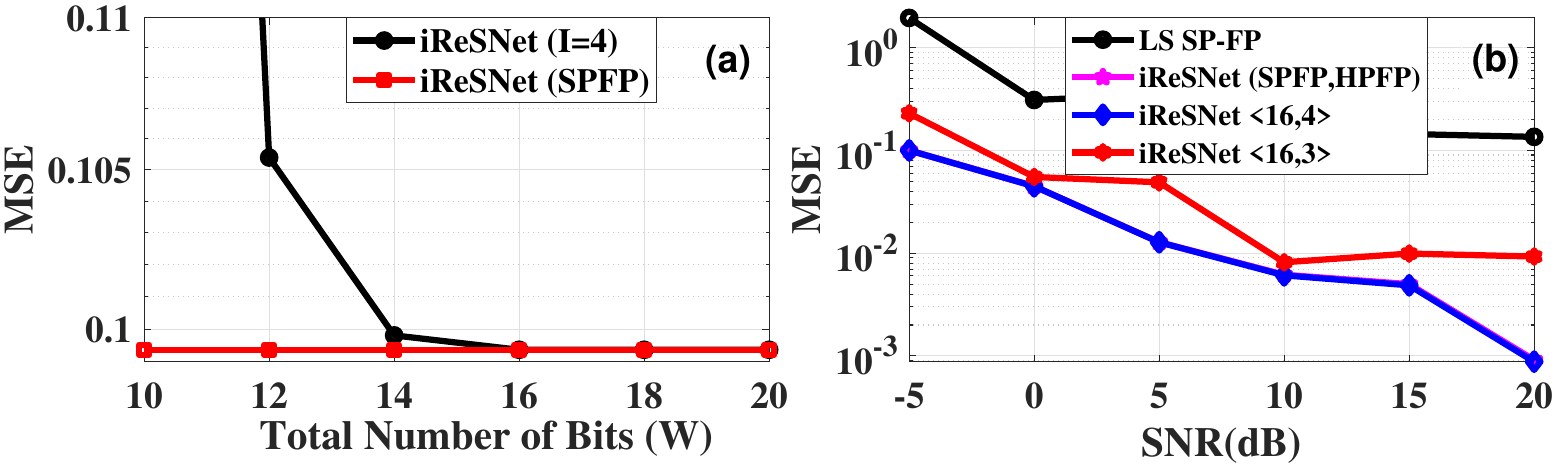}
     \caption{Effect of word length on the MSE performance of iResNet\cite{iResNet} for fixed $I=4$.}
    \label{fig:IReSNet_WLMSE}
\end{figure}

\begin{figure}[!h]
    \centering
    \includegraphics[scale=0.345]{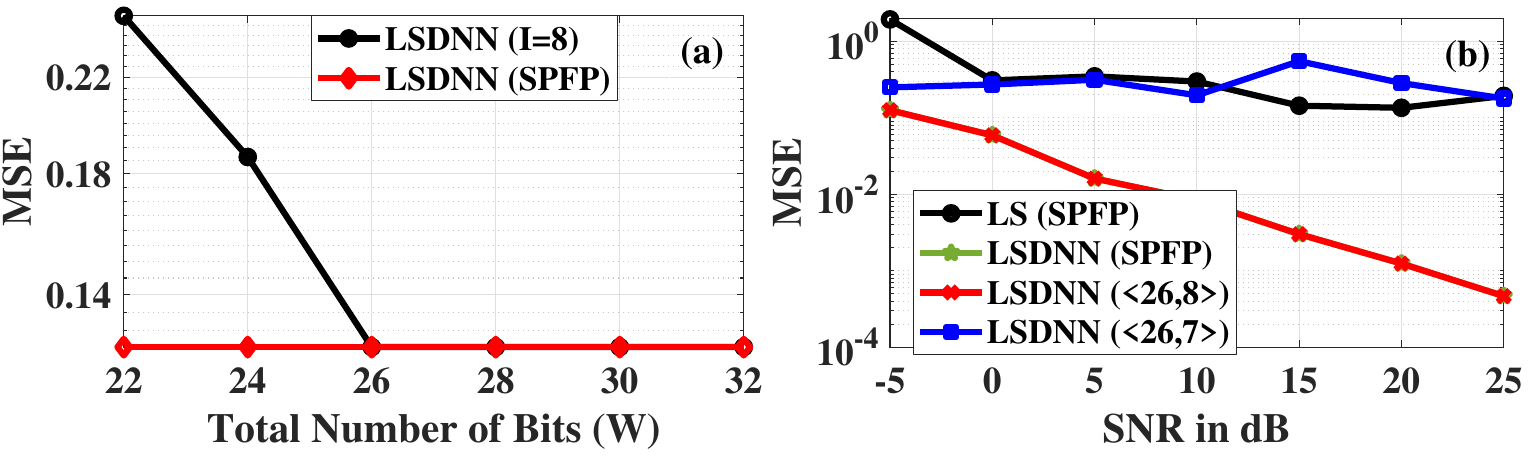}
      \caption{Effect of word length on the MSE performance of LSiDNN for fixed $I=8$.}
    \label{fig:LSDNN_WSel}
\end{figure}

For the selected value of $I$, we find out the minimum value of $W$ that offers the same MSE performance as that of SPFL WL. For LS with $I=4$, we increase the value of $W$ from 6 onwards, as shown in Fig.~\ref{fig:LS_WL}(a). It can be observed that there is no improvement in MSE after $W > 12$, and hence, the WL of $(12,4)$ is selected. Next, we compare the MSE of fixed-point architecture for a wide range of SNR in Fig.~\ref{fig:LS_WL}(b). It can be observed that the WL of $(12,4)$ offers the same MSE as that of SPFL for all SNRs. Single-bit reduction to WL of $(12,3)$ results in significant degradation in MSE, thereby validating the accurate selection of minimum WL.   Similarly, for iReSNet with $I=4$, we compare the MSE for different values of $W$ in Fig.~\ref{fig:IReSNet_WLMSE} (a), and the minimum value of $W$ is 16. Thus, the selected WL for iReSNet is $(16,4)$. In Fig.~\ref{fig:IReSNet_WLMSE} (b), we compare the MSE of iReSNet for a wide range of SNR, and it can be observed that single-bit reduction to WL of $(16,3)$ leads to degradation in the performance. Similarly, as shown in Fig.~\ref{fig:LSDNN_WSel} (a) and (b), the selected WL for LSiDNN is $(26,8)$. We have validated these WLs for different wireless channels, and corresponding results are skipped to avoid repetition of plots.

\begin{figure}[!h]
    \centering
    \includegraphics[scale=0.45]{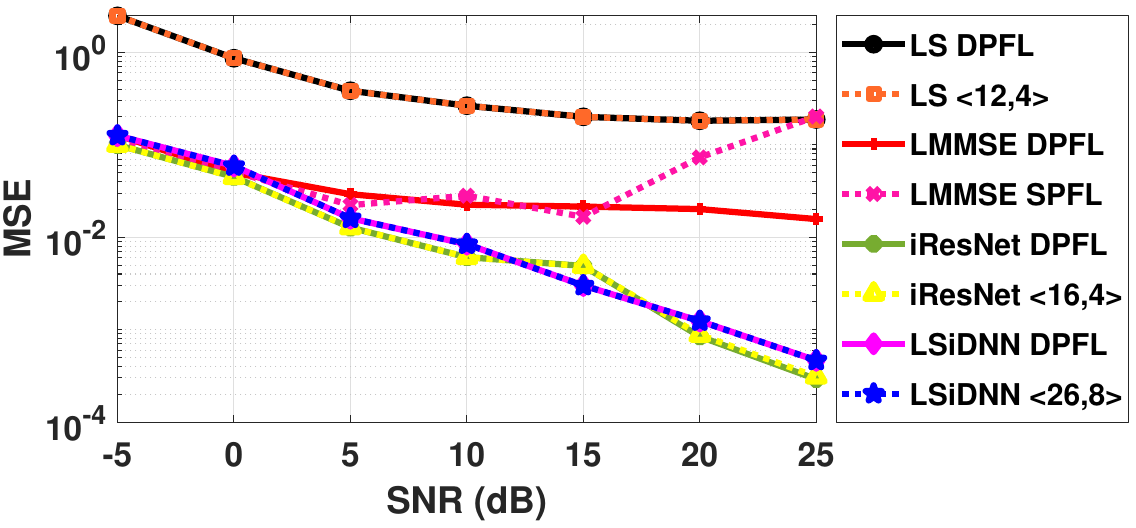}
    \caption{MSE comparison of various optimized architectures on Zynq SoC for EPA channel model.}
    \label{fig:FPCMPALL}
\end{figure}

Next, we compare the performance of all architectures with the minimum selected WL in Fig.~\ref{fig:FPCMPALL} for illustrative EPA  channel. It can be observed that the LMMSE performance degrades significantly for SPFL WL, and hence, fixed-point architecture is not feasible. The fixed-point architectures of DL-based CE offer nearly the same performance as that of DPFL architecture and significantly outperform statistical CE approaches. Similar results are also observed for other wireless channels, and corresponding plots are skipped to avoid repetition of discussion.


\begin{table*}[!h]
    \centering
    \caption{\small Resource utilization and latency comparison of LS, LMMSE, iResNet\cite{iResNet}, and LSiDNN channel estimation for different word-length implementations. }
    \label{tab:Comparison_table}
    \renewcommand{\arraystretch}{1.2}
    \resizebox{\textwidth}{!}{
    \begin{tabular}{@{}|c|c|c|c|c|c|c|c|c|c|@{}}
    \hline
    \textbf{S.No.} & \textbf{Architecture} & \textbf{Word Length} & \textbf{Execution Time (s)} & \textbf{BRAMs} & \textbf{DSPs} & \textbf{LUTs} & \textbf{FFs} & \textbf{Total Power (W)} & \textbf{Dynamic Power (W)} \\ \hline
    \multirow{2}{*}{1} & \multirow{2}{*}{LS} 
      & PL: SPFL & 0.070787 (1.003 x) & 0 & 10 & 8801 & 3703 & 2.220 (1.106 x)& 1.998 (1.117 x)\\ \cline{3-10} 
    &  & PL: (12,4) & 0.070558 (1 x)& 0 & 0 & 2950 & 2462 & 2.209 (1.101 x)& 1.987 (1.111 x)\\ \hline
   \multirow{2}{*}{2} & \multirow{2}{*}{LMMSE} 
      & PL: SPFL & 2.81965 (39.96 x)& 114 & 101 & 25416 & 52564 & 2.340 (1.166 x) & 2.115 (1.182 x)\\ \cline{3-10}
    &  & PL: DPFL & 4.26886 (60.50 x)& 228 & 209 & 25864 & 33402 & 2.931 (1.461 x)& 2.696 (1.507 x)\\\hline
    \multirow{3}{*}{3} & \multirow{3}{*}{IReSNet\_4} & PS: SPFL & 40.25315 (570.49 x)& - & - & - & - & - & -\\ \cline{3-10} 
    &  & PL: SPFL & 18.36422 (260.27 x)& 112 & 100 & 10042 & 9590 & 2.379 (1.185 x) & 2.153 (1.204 x)\\\cline{3-10} 
    &  & PL: (16,4) & 6.63591 (94.04 x)& 56 & 15 & 3956 & 4844 & 2.312 (1.152)& 2.088 (1.167 x) \\ \hline
     \multirow{2}{*}{4} & \multirow{2}{*}{IReSNet\_3} & PS:SPFL & 39.82998 (564.49 x)& - & - & - & - & - & -\\ \cline{3-10}  
    &  & PL (16,4) &  6.50355 (92.17 x)& 56 & 15 & 3956 & 4844 & 2.312 (1.152 x)& 2.088 (1.167 x)\\ \hline   
    \multirow{2}{*}{5} & \multirow{2}{*}{IReSNet\_2} & PS: SPFL &  39.275744 (556.64 x) & - & - & - & - & - & -\\ \cline{3-10}  
    &  & PL: (16,4) & 6.373281 (90.32 x)& 56 & 15 & 3956 & 4844 & 2.312 (1.152 x)& 2.088 (1.167 x)\\ \hline
    \multirow{3}{*}{6} & \multirow{3}{*}{LSiDNN\_LL} & PS: SPFL & 0.76345 (10.82 x)& - & - & - & - & - & -\\ \cline{3-10} 
    &  & PL: SPFL & 0.08784 (1.24 x)& 290 & 40 & 14727 & 10766 & 2.617 (1.304 x)& 2.387 (1.335 x)\\ \cline{3-10} 
    &  & PL: (26,8) & 0.08632 (1.22 x)& 234 & 32 & 10234 & 8561 & 2.605 (1.298 x)& 2.354 (1.316 x)\\ \hline
    \multirow{2}{*}{7} & \multirow{2}{*}{LSiDNN\_CE} & PL:  SPFL & 0.7962 (11.28 x)& 8 & 6 & 2939 & 3543 & 2.006 (1 x)& 1.788 (1 x)\\ \cline{3-10} 
    &  & PL: (26,8) & 0.714885 (10.13 x)& 8 & 6 & 2554 & 3021 & 2.048 (1.020 x)& 1.829 (1.022 x)\\ \hline
    \end{tabular}}
    
    \label{tab:Comparison_table}
\end{table*}

\subsection{ Complexity Analysis}\label{sec:Comparison_Arch}
In this section, we compare the resource utilization, execution time, and power consumption of various CE approaches on the ZSoC platform. 

As shown in Table~\ref{tab:Comparison_table}, we have considered three different architectures of LS. Numerically, LS architecture in PL with WL of (12,4) offers 66\% and 33\% savings in flip-flops (FF) and look-up-table (LUT), respectively, over SPFL architecture. In addition, it eliminates the need for embedded digital signal processing (DSP) units in PL. It can be observed that LMMSE with SPFL and DPFL WLs incur huge resource utilization and power consumption in PL, along with high execution time compared to the rest of the CE architectures. 

Next, we consider three different architectures of iResNet as discussed in Section~\ref{sec:iresnet_arch}. In the case of iResNet\_4, the SPFL architecture in PL offers a 55\% reduction in execution time compared to SPFL architecture in PS for identical functional accuracy. However, this gain in performance is at the additional cost of resource utilization in PL. Using WL of (16,4), iResNet\_4 in PL offers 50\%, 85\%, 60\%, and 49\% reduction in Block RAM (BRAM), Digital Signal Processors (DSP), Flip-Flops (FF), and Lookup Tables (LUT), respectively, for identical functional accuracy. It also offers a 64\% reduction in execution time. This validates the importance of careful WL selection for realization on PL. We have also implemented iResNet\_3 and iResNet\_2 architectures, though both suffer from poor accuracy at high SNR as discussed in  Section~\ref{sec:iresnet_arch}. It can be observed that there is no significant improvement in execution time as well. Hence, iResNet\_4 is the preferred architecture. The iResNet\_4 offers lower complexity and power consumption than LMMSE and better CE performance. However, its latency is almost three times higher than LMMSE. 

Next, we consider two different architectures of LSiDNN as discussed in Section~\ref{sec:lsidnn_arch}. The first architecture, LSiDNN\_LL, is designed to achieve low latency, while the second architecture, LSiDNN\_CE, is designed to achieve low complexity. The LSiDNN\_LL architecture in PL offers almost 86\% reduction in execution time compared to PS, while LSiDNN\_CE in PL offers around 6\% reduction in execution time compared to PS. However, LSiDNN\_CE offers 97\%, 85\%, 80\%, and 67\% reduction in BRAMs, DSP, LUTs, and FFs, respectively, over LSiDNN\_LL. It is interesting to note that the LSiDNN\_LL is the only DL-based CE that offers nearly the same execution time as that of LS with significantly improved CE performance. Thus, depending on the desired latency and resource availability for a given application, any LSiDNN can be explored without compromising functional accuracy.  

Finally, we compare the two DL-based CEs, iResNet\_4 and LSiDNN\_CE. Note that iResNet\_4 is the state-of-the-art DL-based CE in the literature, offering improved performance with low complexity. However, this is the first work where iResNet\_4 has been mapped on the hardware. The proposed LSiDNN\_CE offers an 89\% reduction in execution time with a slight reduction in power consumption. Furthermore, LSiDNN\_CE offers 86\%, 60\%, 36\%, and 37\% reduction in BRAMs, DSP, LUTs, and FFs, respectively, over iResNet\_4. In addition, LSiDNN\_CE offers a 74\% reduction in execution time, 30\% lower power consumption, and 93\%, 94\%, 90\%, and 94\% reduction in BRAMs, DSP, LUTs, and FFs, respectively, over LMMSE.

        

\subsection{Extension to Adaptable Architecture}
\begin{figure}[!h]
    \centering
    \includegraphics[scale=0.51]{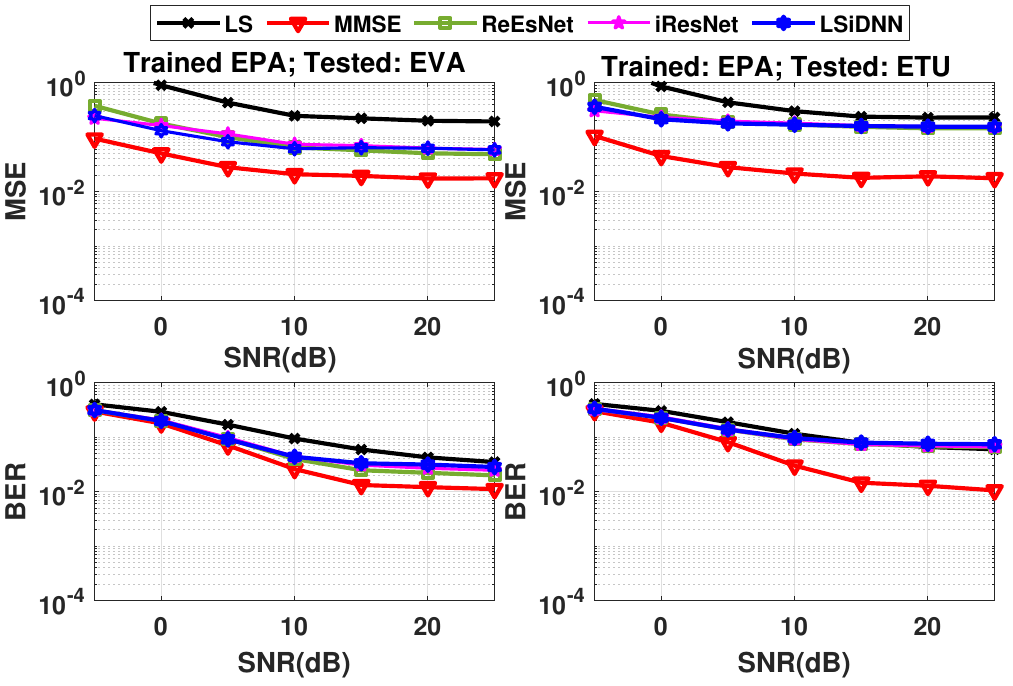}
    \caption{Performance comparison of LSiDNN with other estimation techniques.}
    \label{fig:generazibility}
 \end{figure}

The LS-based CE is popular and widely used in dynamic channel environments compared to LMMSE and DL-based CE, which needs prior knowledge of the channel conditions. Compared to LMMSE, DL-based CE performs well in dynamic mobility conditions. However, as shown in Fig.~\ref{fig:generazibility}, the performance of DL-based CE degrades significantly when testing and training channel conditions do not match \cite{DeepWiPHY2021, haq2022LSDNN}. 
The generalizability problem can be addressed at the algorithmic or system levels. At the algorithmic level, one approach involves designing a generalized DL architecture capable of performing well across various channel conditions. However, such architectures are expected to be computationally complex. Another algorithmic strategy involves online training of DL architectures using the received signals, which demands significant computational capability at edge platforms. Furthermore, the impact of online training-based adaptable architecture on CE latency needs to be studied \cite{Channelformer,onlineLearning2021Mei}. 
A system-level solution involves designing an adaptable architecture capable of reconfiguration based on channel conditions and equipped with intelligence to detect changes in these conditions. Multiple DNN models are trained for various channel conditions, with the system dynamically selecting the most appropriate model as needed. This approach offers computational feasibility and low latency, as the architecture of DNN layers, their sizes, and the number of parameters in each layer remain independent of the channel model in LSiDNN.
Therefore, the same hardware architecture can be reused with different parameters loaded from memory. This is known as memory-based reconfigurability, where trained DNN models are stored in memory, and the parameters corresponding to the current channel conditions are configured on the fly \cite{haq2022LSDNN}. 
Moreover, this approach facilitates site-specific DL-based channel estimation, wherein over-the-air training data is collected for a specific site, and a DL model is trained accordingly. This training can occur offline, and the resulting model can be deployed on-site using reconfigurability.
Though memory-based reconfigurability leads to adaptable architecture, it still needs intelligence to detect the change in channel conditions, for which approaches such as classifier, reinforcement learning, and multi-armed bandit need to be explored. This is the focus of our future work.


\section{Conclusion} \label{conclusion}
In this paper, we studied the feasibility of deep learning (DL) based channel estimation for the wireless physical (PHY) layer on system-on-chip (SoC). We have realized the existing statistical and DL-based channel estimation approaches on SoC via hardware-software co-design and word-length analysis. We show that the existing DL approaches offer improved mean square error and lower bit-error rate than the least square (LS) and linear minimum mean square error (LMMSE) approaches. We observed that DL-based approaches are relatively easy to implement and optimize on FPGA than LMMSE due to simple arithmetic operations. However, they have very high complexity and latency. We designed the LS augmented interpolated deep neural network (LSiDNN) algorithm, which has significantly lower complexity and latency than existing DL approaches for a given MSE and BER performance. Via in-depth experimental results for a wide range of SNR and wireless channels and complexity analysis, we demonstrated the superiority of the proposed LSiDNN approach over existing works. Future works involve the design of intelligent adaptable channel estimation architecture for a dynamic wireless environment.

\appendices
\section{Effect of Training Dataset Size on BER Performance}

\begin{figure}[!htbp]
        \centering
        \includegraphics[scale=0.45]{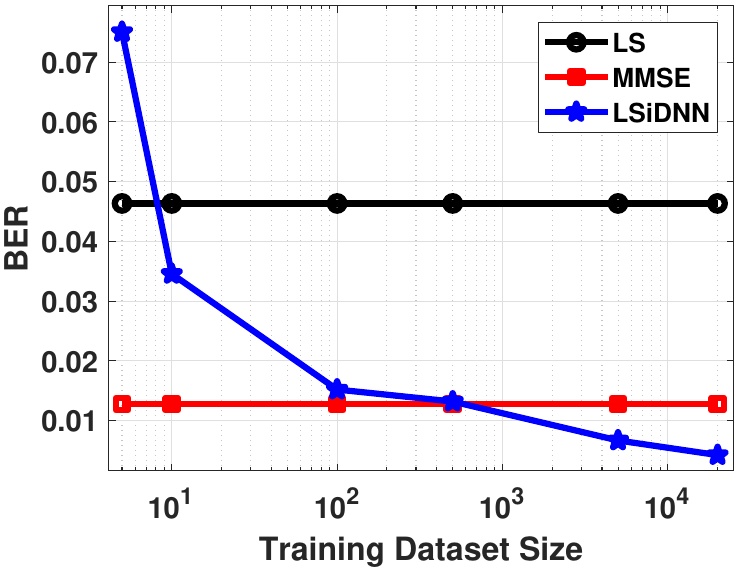}
        \caption{Performance of LSiDNN as a function of dataset Size. }
        \label{fig:datasetSize}
    \end{figure}

The training process in DL depends on the availability of sufficient training data to train the model effectively.
To understand this dependency, we carried out experiments in which we varied the training dataset size, and as shown in Fig.~\ref{fig:datasetSize}, the BER performance of LSiDNN degrades with the decrease in the size of the training dataset. However, the dataset comprising of 10000 data frames, equivalent to 10 seconds of data (1 frame = 1 millisecond), is sufficient for training the LSiDNN, and the availability of such dataset may not be challenging.


\bibliographystyle{IEEEtran}
\bibliography{biblio.bib}
\end{document}